\documentclass{article}
\usepackage{hyperref} 
\usepackage[utf8]{inputenc}
\usepackage[a4paper, total={6in, 10in}]{geometry}

\usepackage[linesnumbered,lined,commentsnumbered]{algorithm2e}
\usepackage[usenames,dvipsnames,svgnames]{xcolor}

\usepackage{algpseudocode}

\usepackage[english]{babel}
\usepackage{amsthm}
\usepackage{color,soul}
\usepackage{appendix}

\usepackage{tikz}  
\usetikzlibrary{positioning}

\usepackage{tablefootnote}
\usepackage{amssymb}

\usepackage{adjustbox}
\usepackage{pifont}% http://ctan.org/pkg/pifont

\newcommand{\cmark}{\ding{51}}%
\newcommand{\xmark}{\ding{55}}%

\newtheorem{theorem}{Theorem}
\newtheorem{conjecture}{Conjecture}
\newtheorem{lemma}{Lemma}
\newtheorem{definition}{Definition}
\newtheorem{example}{Example}

\usepackage{array,etoolbox}
\preto\tabular{\setcounter{magicrownumbers}{0}}
\newcounter{magicrownumbers}
\newcommand\rownumber{\stepcounter{magicrownumbers}\arabic{magicrownumbers}}

\definecolor{colorOne}{rgb}{0.93, 0.18, 0.0}
\definecolor{colorTwo}{rgb}{0.91, 0.84, 0.42}
\definecolor{colorThree}{rgb}{0.74,0.83,1}
\definecolor{colorFour}{rgb}{0.6, 0.84, 0.42}

\definecolor{colorOne}{rgb}{1, 01, 01.0}
\definecolor{colorTwo}{rgb}{01, 01, 01.0}
\definecolor{colorThree}{rgb}{01, 01, 01.0}
\definecolor{colorFour}{rgb}{01, 01, 01.0}

\makeatletter
\renewcommand{\@algocf@capt@plain}{above}% formerly {bottom}
\makeatother



\title{A $1.9\dot{9}$-approximation algorithm for the vertex cover problem on undirected, unweighted graphs(?)}
\title{An Unconditional Deterministic Polynomial-time Algorithm for the Vertex Cover Problem}
\title{Breadth-First Search, Maximal Matching, Local Minimum, and Minimum Vertex Cover}
\title{An Amalgamation of Breadth-First Search, Maximal Matching, and Local Minimum}
\title{On the Amalgamation of Breadth-First Search, Maximal Matching, and Local Minimum}
\title{On the Amalgamation of Breadth-First Search, Maximal Matching and Local Minimization}
\title{Maximum Matching, Breadth-First Search, Maximal Matching and Local Minimization}
\title{On Efficient Election of Diverse Committees}
\title{On Efficient Computation of Diverse Committees}
\title{On Efficient Computation of DiRe Committees}

\author{Kunal Relia\footnote{This work, conceptualized while the author was a student at New York University, was then generously supported in part by Julia Stoyanovich's NSF grants No. 1916647, 1934464, and 1916505. Independent Researcher; kunal.relia91@gmail.com or krelia@nyu.edu.}}
\author{Kunal Relia\footnote{This work, while the author was a student at New York University, was generously supported in part by Julia Stoyanovich's NSF grants No. 1916647, 1934464, and 1916505. Independent Researcher. Correspondence: kunal.relia91@gmail.com or krelia@nyu.edu.}}
\date{October 01, 2025}

\begin{document}

\maketitle

\begin{abstract}

Consider a committee election consisting of (i) a set of candidates who are divided into arbitrary groups each of size \emph{at most} two and a diversity constraint that stipulates the selection of \emph{at least} one candidate from each group and (ii) a set of voters who are divided into arbitrary populations each approving \emph{at most} two candidates and a representation constraint that stipulates the selection of \emph{at least} one candidate from each population who has a non-null set of approved candidates. 

The DiRe (Diverse + Representative) committee feasibility problem  ({a.k.a.} the minimum vertex cover problem on unweighted undirected graphs) concerns the determination of the smallest size committee that satisfies the given constraints. Here, for this problem, we propose an algorithm that is an amalgamation of maximum matching, breadth-first search, maximal matching, and local minimization. We prove the algorithm terminates in polynomial-time. We conjecture the algorithm is an unconditional deterministic polynomial-time algorithm. %A positive resolution of this conjecture would imply resolution of the P vs NP problem, or it can inform further research in this direction. 
\end{abstract}

\clearpage
\tableofcontents
\clearpage

\noindent \emph{\textbf{Preface:} The DiRe committee feasibility problem (stated in the abstract) and the vertex cover problem on unweighted undirected graphs are equivalent (vertices $=$ candidates; edges $=$ candidate groups / voter populations' approved candidates; for details, see Appendix~\ref{sec:dcvcequi}). Hence, for technical simplicity, we henceforth focus the discussion on the latter problem.}

\section{Introduction}
\label{sec:intro}

Given an unweighted undirected graph (specifically, a 2-uniform hypergraph), the vertex cover of the graph is a set of vertices that includes at least one endpoint of every edge of the graph. Formally, given a graph $G = (V, E)$ consisting of a set of vertices $V$ and a collection $E$ of 2-element subsets of $V$ called edges, the vertex cover of the graph $G$ is a subset of vertices $S \subseteq V$ that includes at least one endpoint of every edge of the graph, i.e., for all $e \in E$, $e\cap S \neq \phi$. The corresponding computational problem of finding the minimum-size vertex cover (MVC) is NP-complete\footnote{Strictly speaking, the decision version of the vertex cover problem is NP-complete whereas MVC itself (search version) is NP-hard. See Section 2.1 of \cite{khot2019proof} for a lucid explanation delineating (a) search and decision problems and (b) NP-hardness and NP-completeness.} \cite{cook1971complexity,levin1973universal,karp1972reducibility}, which means that there is no \emph{known} deterministic polynomial-time algorithm to solve MVC. Here, we conjecture an unconditional deterministic polynomial-time algorithm for MVC on unweighted simple connected graphs\footnote{We subtly yet drastically switch the discussion from unweighted undirected graphs to unweighted simple connected graphs. For simplicity, we want to avoid having loops and/or unconnected components in the graph. In the context of this paper, this switch has no impact on the NP-completeness of the problem (Appendix~\ref{sec:vcuscnpc}). {Notwithstanding, in the case of the presence of loops, our algorithm will work (with minor modifications) if each loop is replaced by adding a dummy vertex and a corresponding edge. In the case of unconnected components, we can run the algorithm for each connected component independently and take a union of each of the minimum vertex covers to get the final minimum vertex cover.}}.

\noindent\marginbox{0pt 0pt 0pt 10pt}{\fcolorbox{black}{lightgray!75}{\vbox{\emph{We sparingly use ``Non-technical Comment'' boxes in this paper. These comments are not a part of the paper in a technical sense but they provide important answers to some non-technical but important ``whys'' and ``so whats'' of the paper. It may help a reader relate to the journey of working on the paper. \\\\\textbf{Non-technical Comment:} A chance re-encounter with one of Aesop's fables, ``The Fox and the Grapes'', from my childhood days was a motivation to begin thinking about this paper. By calling the DiRe committee feasibility problem ``hard'' (NP-hard), was I being the fox who found the grapes sour
%instead of my inability to find an efficient algorithm to reduce inequality
?}}}}
\section{Notation and Preliminaries}
\label{sec:notations}

We now formally define the computation problems related to finding the vertex cover of a given graph. First, we define the search/optimization problem:

\begin{definition}[Minimum Vertex Cover Problem (MVC)]\label{def:MVCsearch}
Given a graph $G$, what is the smallest non-negative integer $k$ such that the graph $G$ has a vertex cover $S$ of size $k$?
\end{definition}

Next, we restate the above as a decision problem:
\begin{definition}[Vertex Cover Problem (VC)]\label{def:MVCdecision}
Given a graph $G$ and a non-negative integer $k$, does the graph $G$ have a vertex cover $S$ of size at most $k$?
\end{definition}
Unless stated otherwise, we henceforth discuss solving VC (i.e. Definition~\ref{def:MVCdecision}), which is actually NP-complete.
\section{Algorithm Overview}
\label{sec:algo/overview}

The algorithm is broadly divided into four phases. The first three phases are (slightly adapted versions of) algorithms for three known problems, namely maximum matching, breadth-first search, and maximal matching. The last phase is a technique we call local minimization. We now discuss these phases and give an overview of the algorithm.

\begin{definition}[Matching]\label{def:M} 
Given a graph $G$, a matching $M$ is a subset of the edges $E$ such that no vertex $v \in V$ is incident to more that one edge in $M$.
\end{definition}
Alternatively, we can say that given a graph $G$, no two edges in a matching $M$ have a common vertex.

\subsection{Maximum Matching}

Phase 1 of the algorithm finds maximum matching of the input graph:

\begin{definition}[Maximum Matching]\label{def:MM}
Given a graph $G$, a matching $M$ is said to be maximum if for all other matching $M'$, $|M| \geq |M'|$.
\end{definition}

Equivalently, the size of the maximum matching $M$ is the (co-)largest among all the matching. Next, there is a known relationship between the size of maximum matching and the size of minimum vertex cover: 
\begin{lemma}\label{lemma:MaximumMatchingVertexCover}
In a given graph $G$, if $M$ is a maximum matching and $S$ is a minimum vertex cover, then $|S|\geq|M|$.
\end{lemma}

Lemma~\ref{lemma:MaximumMatchingVertexCover} means that the largest number of edges in a matching does not exceed the smallest number of vertices in a cover. We use this fact to set a lower bound on the size of the minimum vertex cover and terminate the algorithm early if the integer $k$ is less than $|M|$. %This is because we are certain that we can not get a smaller vertex cover. 

\subsection{Breadth-first Search}
Phase 2 of the algorithm stores the vertices at each level of the tree derived using breadth-first search (BFS):
\begin{definition}[Breadth-First Search]\label{def:BFS}
Given a graph $G$, a Breadth-first Search (BFS) algorithm seeds on a root vertex $v \in V$ and visits all vertices at the current depth level of one. Then, it visits all the nodes at the next depth level. This is repeated until all vertices are visited.
\end{definition}

While the BFS algorithm is canonically a search algorithm, we use it here to derive a tree. This tree itself is not needed. Only the information of the level at which each vertex is in the tree is stored for use during the third phase. 

\subsection{Maximal Matching}
Phase 3 of the algorithm entails the use of maximal matching. 

\begin{definition}[Maximal Matching]\label{def:MaM}
Given a graph $G$, a matching $M$ is said to be maximal if for all other matching $M'$, $M \not \subset M'$. 
\end{definition}

In other words, a matching $M$ is maximal if we cannot add any new edge $e \in E$ to the existing matching. During this maximal matching phase, the edges are selected using a specific procedure that uses  information stored (i) regarding the edges that are a part of the maximum  matching and (ii) about the vertices present at each level of the tree derived using BFS. Additionally, during each iteration of maximal matching, the algorithm stores the \emph{current} neighboring vertices of each endpoint. We call this as an endpoint vertex \emph{representing} its neighboring vertex. 

\begin{definition}[Represents\footnote{The term is inspired by a type of multiwinner election where the aim is to elect the smallest committee that represents every voter. In our context, we want to select the smallest set of vertices that covers (represents) each edge.}]\label{def:representing}
Given a graph $G$, a vertex $u \in V$ is said to \textbf{represent} a vertex $v \in V$ when vertex $v$ is currently connected to vertex $u$ by an edge $e \in E$. Conversely, vertex $v$ is \textbf{represented by} vertex $u$.
\end{definition}

Observe that when some vertex $u$ \emph{currently} represents a vertex $v$, the algorithm is essentially storing information about the presence of an edge connecting the two vertices. There is stress on the word currently as for a given iteration, an edge should not have been removed. The information is stored in \emph{represents table} that consist of \emph{represents lists.}

\begin{definition}[Represents Table]\label{def:representTable}
A represents table $R$ is a 2-column table that stores the endpoints of edges selected during maximal matching and the vertices each endpoint represents. 
\end{definition}

\begin{definition}[Represents List]\label{def:representList}
Given a represents table $R$, a vertex $u \in V$  that is represented by a vertex $v \in V$ is said to be in the represents list of $v$.
\end{definition}

Finally, in the last step of an iteration of the maximal matching phase, the algorithm removes the edge that connects (i) the two endpoints and (ii) endpoints and their respective neighbors. 

\begin{example}\label{eg:example}
Consider the following graph $G$:
\begin{center}
    \begin{tikzpicture}  
      [scale=.9,auto=center,every node/.style={circle,fill=blue!20}] % here, node/.style is the style pre-defined, that will be the default layout of all the nodes. You can also create different forms for different nodes.  
        
      \node (a1) at (1,1) {0};  
      \node (a2) at (3,1)  {1}; % These all are the points where we want to locate the vertices. You can create your diagram first on a rough paper or graph paper; then, through the points, you can create the layout. Through the use of paper, it will be effortless for you to draw the diagram on Latex.  
      \node (a3) at (5,1)  {2}; 
      \node (a4) at (7,1)  {3}; 
      
      \draw (a1) -- (a2); % these are the straight lines from one vertex to another  
      \draw (a2) -- (a3); 
      \draw (a3) -- (a4); 
      
    \end{tikzpicture}  
\end{center}
During maximal matching, assume that the algorithm first selects the edge connecting vertex 0 and vertex 1. Then, the endpoints of the selected edge are 0 and 1. For each endpoint, the algorithm stores the information of the vertices it represents. Here, vertex 0 represents \{1\} and vertex 1 represents \{0, 2\}. All the edges connected  to the two endpoints in any way are removed. 
\begin{center}
    \begin{tikzpicture}  
      [scale=.9,auto=center,every node/.style={circle,fill=blue!20}] % here, node/.style is the style pre-defined, that will be the default layout of all the nodes. You can also create different forms for different nodes.  
        
      \node (a1) at (1,1) {0};  
      \node (a2) at (3,1)  {1}; % These all are the points where we want to locate the vertices. You can create your diagram first on a rough paper or graph paper; then, through the points, you can create the layout. Through the use of paper, it will be effortless for you to draw the diagram on Latex.  
      \node (a3) at (5,1)  {2}; 
      \node (a4) at (7,1)  {3}; 
      
      %\draw (a1) -- (a2); % these are the straight lines from one vertex to another  
      %\draw (a2) -- (a3); 
      \draw (a3) -- (a4); 
      
    \end{tikzpicture} 
\end{center}
In the next iteration of maximal matching, the algorithm selects the edge connecting vertex 2 and vertex 3. The two endpoints represent each other only. Specifically, vertex 2 represents \{3\} and vertex 3 represents \{2\}. All the edges connected  to the two endpoints in any way are removed. 
\begin{center}
    \begin{tikzpicture}  
      [scale=.9,auto=center,every node/.style={circle,fill=blue!20}] % here, node/.style is the style pre-defined, that will be the default layout of all the nodes. You can also create different forms for different nodes.  
        
      \node (a1) at (1,1) {0};  
      \node (a2) at (3,1)  {1}; % These all are the points where we want to locate the vertices. You can create your diagram first on a rough paper or graph paper; then, through the points, you can create the layout. Through the use of paper, it will be effortless for you to draw the diagram on Latex.  
      \node (a3) at (5,1)  {2}; 
      \node (a4) at (7,1)  {3}; 
      
      %\draw (a1) -- (a2); % these are the straight lines from one vertex to another  
      %\draw (a2) -- (a3); 
      %\draw (a3) -- (a4); 
      
    \end{tikzpicture}  
\end{center}
Finally, the following information is stored by the algorithm:
\\
\begin{table}[h]
    \centering
    \begin{tabular}{l|l}
        Node 1 & Node 2\\
        \hline
        0 - \{1\} & \colorbox{blue!20}{1 - \{0, 2\}}\\
        2 - \{3\} & 3 - \{2\} \\
    \end{tabular}
    \caption{Information stored in a \textbf{``Represents Table''} $R$ after the end of maximal matching phase.}
    \label{tab:maximalmatchingexample}
\end{table}
\\
The information contained in row 1 under ``Node 2'' of Table~\ref{tab:maximalmatchingexample} is: vertex 1 is an endpoint vertex that represents vertices 0 and 2. Conversely, vertices 0 and 2 are represented by endpoint vertex 1. Also, vertices 0 and 2 are in the \textbf{represents list} of endpoint vertex 1.
\end{example}

Two known facts related to maximal matching will be useful later:

\begin{lemma}\label{lemma:MaMisVC}
The endpoints of a maximal matching form a vertex cover.
\end{lemma}

\begin{lemma}\label{lemma:MMisMaM}
In a graph $G$, if a matching $M$ is maximum, it implies the matching $M$ is also maximal. The converse does not hold.
\end{lemma}

%The converse of Observation~\ref{lemma:MMisMaM} does not hold. 
%We specifically use Lemma~\ref{lemma:MMisMaM} in Section~\ref{sec:proof} and explain why the third phase is called maximal matching and not maximum matching. 

We may use Lemma~\ref{lemma:MMisMaM} in the proof of correctness and explain why the third phase is called maximal matching and not maximum matching.

\subsection{Local Minimization}
The last Phase, local minimization, is a new technique. It is not adapted from any known techniques to the best of our knowledge. Also, note that our version of local minimization is not related to the local search used in heuristic algorithms. We use the term \emph{local} in local minimization because the vertex cover we get at the end of this phase is the ``smallest'' and not necessarily minimum. Specifically, the vertex cover we get is dependent on the endpoints of the edges selected during the maximal matching phase. Hence, from a given set of vertices, local minimization phase uses three stages to select a vertex cover of the smallest possible size, which may not be the minimum vertex cover: 
\begin{enumerate}
    \item \textbf{Freeze ``necessary'' vertices:} Freeze each endpoint $v$ in the represents table $R$ that represents a vertex $u$ that is not an endpoint in $R$. Vertex $u$ can not be in the vertex cover $S$ as it is not an endpoint of any edge selected during maximal matching. Hence, vertex $v$ necessarily needs to be a part of the vertex cover to cover the edge connecting $u$ and $v$. 
    \item \textbf{Top-down removal of ``terminal'' vertices:} Remove each endpoint with degree one in graph $G$. The other endpoint is simultaneously frozen. 
    \item \textbf{Bottom-up freeze and remove:} Freeze and remove ``necessary'' and ``terminal'' vertices, respectively, based on the \emph{current} state of table $R$. 
\end{enumerate}

\begin{definition}[Local Minimization]\label{def:LM}
Given a graph $G$, a subset of vertices $V' \subseteq V$ that covers all edges and for each vertex $v \in V'$ the list of vertices it represents, the local minimization selects the smallest sized subset of vertices $S' \subseteq V'$ such that each edge is covered.  
\end{definition}

% \begin{enumerate}
%     \item breadth-first search (BFS): Execute BFS using your favorite polynomial-time algorithm. Ties are broken and vertices at the same level are sorted using lexicographic ordering\footnote{Henceforth, all ties are broken and all ordering (sorting) is done based on lexicographic ordering unless noted otherwise. The ordering does not impact the correctness but ensures that for same input, the output remains exactly the same.}. 
%     \item maximal matching: Maximal matching implies a vertex cover.
%     \item local minimization:
% \end{enumerate}

\subsection{Summary}
The algorithm we discovered is an amalgamation of the above-discussed phases. The sequential implementation of these phases ensures we get a minimum vertex cover cover. At a high-level, this is because: (i) Maximum matching and breadth-first search ensures that the edges selecting during the maximal matching phase follows a procedure as opposed to vanilla maximal matching where edges are selected randomly. (ii) Maximal matching implies we get a vertex cover. (iii) Local minimization ensures we get the smallest vertex cover. Overall, we conjecture that the combination of all these implies we get the minimum vertex cover. 

\noindent\marginbox{0pt 0pt 0pt 10pt}{\fcolorbox{black}{lightgray!75}{\vbox{\emph{\\\textbf{Non-technical Comment:} As discussed, the flow of the algorithm is as follows: maximum matching $\rightarrow$ breadth-first search $\rightarrow$ maximal matching $\rightarrow$ local minimization. However, the evolution of the algorithm happened in the following order:
maximal matching $\rightarrow$ local minimization $\rightarrow$ breadth-first search $\rightarrow$ maximum matching. Indeed, eventually ``prefixing'' the algorithm with maximum matching helped us deal with the messy cycles, especially odd cycles. Recall that Blossom algorithm \cite{edmonds1965paths} had to do ``extra work'' just to deal with odd cycles.}}}}
\section{Algorithm}\label{sec:algo}

We now present the core contribution of this paper, an algorithm to solve the VC problem. 
%For presentation purposes, we have color-coded\footnote{The algorithm is not dependent on the color-coding and can be interpreted independently. Thus, this paper remains inclusive to every eye-sight. The colors enhance the interpretation.} the four phases that make up the algorithm:
% \begin{itemize}
%     \item Maximum matching is coded \colorbox{colorOne}{red}
%     \item Breadth-first Search is coded \colorbox{colorTwo}{yellow}
%     \item Maximal matching is coded \colorbox{colorThree}{blue}
%     \item Local minimization is coded \colorbox{colorFour}{green}
% \end{itemize}
%Please note that the color coding is approximate and due to the presence of loops, one phase, in principle, trickles over to the next one. Also, henceforth, 
In the algorithm, all ties are broken and all ordering (sorting) of vertices is done based on lexicographic ordering unless noted otherwise. The ordering does not impact the correctness but ensures that for same input, the output remains the same.

\begin{center}
        \begin{minipage}{\textwidth}
            \SetAlgoLined
            \SetNlSty{textbf}{}{:}

            \begin{algorithm}[H]
                \caption{$\mathtt{Vertex\_Cover}$($G$, $k$)}
                \label{alg:VC}
            
                \DontPrintSemicolon
                
                \KwData{Graph $G$ = $(V, E)$, non-negative integer $k$}
                \KwResult{returns ``YES'' if there is a vertex cover $S$ of size at most $k$, ``NO'' otherwise }
                
                \vspace{0.25cm}

                %$S$ = $V$\\
                $V_s$ = lexicographically sorted vertices\\
                %$R$ = a two-column table that stores the endpoints of an edge picked during maximal matching and the corresponding vertices each endpoint represents\\
                \colorbox{colorOne}{$E_{M}$ = maximum matching found using the Blossom Algorithm \cite{edmonds1965paths}}\\
                \If{$k < |E_M|$}
                    {return ``NO''}
                \For{\normalfont{\textbf{each }}\colorbox{colorTwo}{$v$ $\in$ $V_s$}}{
                
                    \colorbox{colorTwo}{$BFS_{level}$ = an array of arrays storing sorted vertices at each level of} 
                    \colorbox{colorTwo}{breadth-first search tree seeded on $v$}\\
                    \colorbox{colorThree}{$R$ = $\mathtt{Maximal\_Matching}$($G$, $E_M$, $BFS_{level}$)}\\
                    \colorbox{colorFour}{$S$ = $\mathtt{Local\_Minimization}$($R$)}\\
                    \If{$|S|\leq k$}
                    {return ``YES''}
                    %if $|S| = |E_M|$ then return ``NO''\\
                }
                return ``NO''\\
            \end{algorithm}
      \end{minipage}
\end{center}

\noindent\marginbox{0pt 0pt 0pt 10pt}{\fcolorbox{black}{lightgray!75}{\vbox{\emph{\\\textbf{Non-technical Comment:} The technical discussion for each of the phases of the algorithm will follow in the succeeding sections. Here, we share our non-technical motivation for including maximum matching and BFS phases in the algorithm. Our guiding question was ``Is it possible that we have missed out on considering all the factors that decide the vertices being selected to form the minimum-size vertex cover?'' Such factors may not be given to us in the traditional sense and hence, may not be ``visible''. We may have to infer them to use them. We do so in this paper. Given an unweighted undirected graph for VC problem, maximum matching and BFS lend inherent edge weights and directions, respectively. After traversing through the algorithm, it will be intuitively evident that during maximal matching, each edge carries certain ``weight'' and the  edge selections happen in particular ``direction''. Thus, identifying and including such factors was another motivation for this paper. 
 }}}}

\begin{center}
        \begin{minipage}{\textwidth}
            \SetAlgoLined
            \SetNlSty{textbf}{}{:}

            \begin{algorithm}[H]
                \caption{$\mathtt{Maximal\_Matching}$($G$, $E_M$, $BFS_{level}$)}
                \label{alg:MaM}
            
                \DontPrintSemicolon
                
                \KwData{Graph $G$ = $(V, E)$, Edges in maximum matching $E_M$, Levels at which each vertex is present after BFS $BFS_{level}$ }
                \KwResult{returns $R$ - Represents Table}
                
                \vspace{0.25cm}

                $R$ = a two-column table, Represents Table, that stores the endpoints of an edge selected during maximal matching and the corresponding vertices each endpoint represents\\
                \For{\normalfont{\textbf{each }}$level$ in $BFS_{level}$}{
                    \While{there is an unvisited vertex in level}{

                        \uIf{there exists an edge that connects two vertices on the same level and is in $E_M$}
                            {select the edge}
                        \uElseIf{there exists an edge that connects two vertices on the same level and is not in $E_M$}
                            {select the edge}
                        \uElseIf{there exists an edge that connects one vertex on the current level with another vertex on the next level and is in $E_M$}
                            {select the edge}
                        %\ElseIf{there exists an edge that connects one vertex on the current level with another vertex on the next level and is not in $E_M$}
                        %     {select the edge}
                        \Else{select the edge that connects one vertex on the current level with another vertex on the next level and is not in $E_M$}
                        % \colorbox{colorThree}{select an edge using the following order of precedence:}\\
                        %     \colorbox{colorThree}{(a) an edge that connects two vertices on the same level and }
                        %     \colorbox{colorThree}{is in $E_M$ }\\
                        %     \colorbox{colorThree}{(b) an edge that connects two vertices on the same level and }
                        %     \colorbox{colorThree}{is not in $E_M$ }\\
                        %     \colorbox{colorThree}{(c) an edge that connects one vertex on the current level with }
                        %     \colorbox{colorThree}{another vertex on the next level and is in $E_M$ }\\
                        %     \colorbox{colorThree}{(d) an edge that connects one vertex on the current level with }
                        %     \colorbox{colorThree}{another vertex on the next level and is not in $E_M$ } \\ 
                        %\vspace{0.2cm}
                        Mark the two endpoints of the selected edge as visited in $BFS_{level}$\\
                        Append after the last row of $R$ the two endpoints of the selected edge and the respective vertices each endpoint represents\\
                        Remove from graph $G$ the selected edge and all the edges that are connected to the two endpoints\\
                        If any vertex becomes edgeless in $G$, mark the vertex as visited in $BFS_{level}$\\
                    }
                }
                   
                return $R$\\
            \end{algorithm}
      \end{minipage}
\end{center}

\begin{center}
        \begin{minipage}{\textwidth}
            \SetAlgoLined
            \SetNlSty{textbf}{}{:}

            \begin{algorithm}[H]
                \caption{$\mathtt{Local\_Minimization}$($R$)}
                \label{alg:LM}

                \DontPrintSemicolon

                \KwData{Represents Table $R$}
                \KwResult{returns $S$ - the smallest vertex cover}
                
                \vspace{0.25cm}

                $S$ = $\phi$ \\
                $P$ = set of endpoints in $R$ selected during maximal matching\\
                \For{\normalfont{\textbf{each }}endpoint vertex $v$ in $R$}{
                    \If{$v$ represents at least one vertex not in $P$}{
                        //freeze vertex $v$ but do not remove any vertex from $R$\\
                        $R$, $S$ = $\mathtt{Freeze\_and\_Remove}$($R$, $S$, $v$, $\phi$)\\
                    }
                }
                %\colorbox{colorFour}{For each vertex $s$ in $S$, remove $s$ from every represents list in $R$}\\
                // The following for loop will traverse through the table $R$ top-down\\
                \For{\normalfont{\textbf{each }}$row$ in $R$}{
                    \uIf{if any one endpoint in $row$ is either frozen or removed}{
                        \textbf{continue}
                    }
                    \ElseIf{one endpoint $u$ in $row$ only represents another endpoint vertex $v$ in $row$ and $v$ represents more than one vertex}{
                        \If{$u$ is not represented by any endpoint in $R$ other than $v$}{
                            %//remove $u$ as it has degree one in input graph $G$\\
                            $R$, $S$ = $\mathtt{Freeze\_and\_Remove}$($R$, $S$, $v$, $u$)\\
                        }
                    }
                }
                // The following for loop will traverse through the table $R$ bottom-up\\
                \For{\normalfont{\textbf{each }}$row$ in $R$}{
                    \uIf{(if both endpoints are frozen) or (one endpoint is frozen and one is removed)}{
                        \textbf{continue}
                    }
                    \uElseIf{endpoint $u$ remains and endpoint $v$ is removed}{
                        $R$, $S$ = $\mathtt{Freeze\_and\_Remove}$($R$, $S$, $u$, $\phi$)\\
                    }
                    
                    %\ElseIf{both endpoints $u$ and $v$ in $row$ represent exactly one vertex}{
                    \Else{ 
                    //at this point, both endpoints $u$ and $v$ in $row$ represent exactly one vertex, namely each other\\
                        \uIf{$u$ is represented by more endpoints in $R$ than $v$}{
                             $R$, $S$ = $\mathtt{Freeze\_and\_Remove}$($R$, $S$, $u$, $v$)\\
                        }
                        \uElseIf{$v$ is represented by more endpoints in $R$ than $u$}{
                            $R$, $S$ = $\mathtt{Freeze\_and\_Remove}$($R$, $S$, $v$, $u$)\\
                        }
                        \Else{
                            $R$, $S$ = $\mathtt{Freeze\_and\_Remove}$($R$, $S$, $u$, $v$)\\
                        }
                    }
                }
                   
                return $S$\\
            \end{algorithm}
      \end{minipage}
\end{center}

\begin{center}
        \begin{minipage}{\textwidth}
            \SetAlgoLined
            \SetNlSty{textbf}{}{:}

            \begin{algorithm}[H]
                \caption{$\mathtt{Freeze\_and\_Remove}$($R$, $S$, $freeze$, $remove$)}
                \label{alg:FnR}
            
                \DontPrintSemicolon
                
                \KwData{Represents Table $R$, Vertex Cover $S$, vertex to be frozen $freeze$, vertex to be removed $remove$}
                \KwResult{returns Represents Table $R$, Vertex Cover $S$}
                
                \vspace{0.25cm}
                Remove vertex $remove$ and its represents list from $R$\\
                Freeze vertex $freeze$ in $R$\\
                Append vertex $freeze$ to $S$\\
                Remove vertex $freeze$ from every represents list in $R$\\
                Remove the represents list of vertex $freeze$ in $R$\\

                \For{\normalfont{\textbf{each }}non-frozen and unremoved $endpoint$ in $R$ that represents $remove$}{
                    $R$, $S$ = $\mathtt{Freeze\_and\_Remove}$($R$, $S$, $endpoint$, $\phi$)
                }

                \For{\normalfont{\textbf{each }}non-frozen and unremoved $endpoint$ in $R$ that does not represent any vertex}{
                    $R$, $S$ = $\mathtt{Freeze\_and\_Remove}$($R$, $S$, $\phi$, $endpoint$)
                }
                   
                return $R$, $S$\\
            \end{algorithm}
      \end{minipage}
\end{center}

We conjecture the following: 

\begin{conjecture}\label{thm:main}
    Algorithm~\ref{alg:VC} returns ``Yes'' if and only if a given instance of VC is a ``Yes'' instance.
\end{conjecture}
\section{Time Complexity Analysis}
\label{sec:time}

In this section, we discuss the time complexity of the algorithm (Table~\ref{tab:time_complexityAlg1}, Table~\ref{tab:time_complexityAlg2}, Table~\ref{tab:time_complexityAlg3}, Table~\ref{tab:time_complexityAlg4}). $m$ denotes the number of vertices $V$ and $n$ ($\leq m^2$) denotes the number of edges $E$. 

In each table, we give the complexity of each line (each operation), the complexity of the loop (complexity of line multiplied by the number of loop iterations) and the dominant complexity. For convenience, the beginning of a loop, specifically the number of loop iterations, is highlighted (e.g., \colorbox{blue!20}{Line 6} in Table~\ref{tab:time_complexityAlg1}). Each statement within the loop is prefixed with a pointer ($\blacktriangleright$). In the case of nested loops, an additional pointer ($\rhd$) is used. 

\paragraph{Time complexity of Algorithm~\ref{alg:FnR}:} 
We elaborate upon the time complexity of Algorithm~\ref{alg:FnR} because the time complexity of the remainder of the algorithms is self-explanatory from the respective tables. In Algorithm~\ref{alg:FnR}, we have recursive calls (line 7 and line 10). However, by design, Algorithm~\ref{alg:FnR} can be called at most $m$ times only. This is because each time it is called, at least one vertex is either removed or frozen. Hence, after at most $m$ calls, no unfrozen or unremoved vertex will exist. Each call takes $\mathcal{O}(m^2)$ time. Overall, in the worst case, the height of the recursion tree is $m$ and each level has one subproblem taking  $\mathcal{O}(m^2)$. Thus, total complexity is $\mathcal{O}(m)\cdot$$\mathcal{O}(m^2)=$$\mathcal{O}(m^3)$.

\begin{theorem}\label{thm:time}
The asymptotic running time of Algorithm~\ref{alg:VC} is $\mathcal{O}(m^3n^2)$.
\end{theorem}

\begin{proof}
    Line 8 in Algorithm~\ref{alg:VC}  dominates 
    %or is asymptotically equivalent to 
    the complexity of all other lines  as shown in Table~\ref{tab:time_complexityAlg1}. This dominant complexity is $\mathcal{O}(m^3n^2)$. Hence, the time complexity of the entire algorithm is $\mathcal{O}(m^3n^2)$.
\end{proof}

On one hand, asymptotically, $\mathcal{O}(n)$ = $\mathcal{O}(m^2)$. This is because the maximum number of edges ($n$) possible in a simple graph is $\frac{m\cdot(m-1)}{2}$, which is less than $m^2$. On the other hand, asymptotically, $\mathcal{O}(n)$ = $\mathcal{O}(m)$. This is because the minimum number of edges ($n$) needed in a connected graph is $m-1$. In either case, the dominating time complexity discussed in Table~\ref{tab:time_complexityAlg1} remains the same. In the worst case, it dominates the time complexity of all lines. In the case of a sparse graph, it either dominates or is equivalent to  the time complexity of other lines. Hence, the time complexity stated in Theorem~\ref{thm:time} holds.

%alg 2 dominates hewnce taken liberty in 3 and 4. with careful extra work, it can be shown that 3 and 4 are more efficeint. but it does not matter. 

\begin{table}
    \centering
    \begin{tabular}{|c||l|l|l|}
        \hline
        Line  Number& Line  complexity& Loop  complexity& Dominant complexity\\
        % Number &  complexity &  complexity &  complexity\\
        \hline
        \hline
        \rownumber & $\mathcal{O}(m\cdot \log m)$ & - & $\mathcal{O}(m\cdot \log m)$\\
        \rownumber & $\mathcal{O}(m^2n)$ & - & $\mathcal{O}(m^2n)$\\
        \rownumber & $\mathcal{O}(1)$ & - & $\mathcal{O}(m^2n)$\\
        \rownumber & $\mathcal{O}(1)$ & - & $\mathcal{O}(m^2n)$\\
        \rownumber & - & - & $\mathcal{O}(m^2n)$\\
        \rownumber & $\mathcal{O}(1)$ & \colorbox{blue!20}{$\mathcal{O}(m)$} & $\mathcal{O}(m^2n)$\\
        \rownumber & $\mathcal{O}(m+n)$ & $\blacktriangleright$ $\mathcal{O}(m^2+mn)$ & $\mathcal{O}(m^2n)$\\
        \rownumber & $\mathcal{O}(m^2n^2)$ [Table~\ref{tab:time_complexityAlg2}] & $\blacktriangleright$ $\mathcal{O}(m^3n^2)$ & $\mathcal{O}(m^3n^2)$ = $\mathcal{O}(m^7)$\\
        \rownumber & $\mathcal{O}(m^4)$ [Table~\ref{tab:time_complexityAlg3}] & $\blacktriangleright$ $\mathcal{O}(m^5)$ & $\mathcal{O}(m^3n^2)$\\
        \rownumber & $\mathcal{O}(1)$ & $\blacktriangleright$ $\mathcal{O}(m)$ & $\mathcal{O}(m^3n^2)$\\
        \rownumber & $\mathcal{O}(1)$ & $\blacktriangleright$ $\mathcal{O}(m)$ & $\mathcal{O}(m^3n^2)$\\
        \rownumber & - & - & $\mathcal{O}(m^3n^2)$\\
        \rownumber & - & - & $\mathcal{O}(m^3n^2)$\\
        \rownumber & $\mathcal{O}(1)$ & - & $\mathcal{O}(m^3n^2)$\\
        
        % \rownumber & $\mathcal{O}(1)$ & $\blacktriangleright$ \colorbox{blue!20}{$\mathcal{O}(m^2)$} & $\mathcal{O}(m^2n)$\\
        % \rownumber & $\mathcal{O}(m\cdot \log m)$\tablefootnote{For simplicity, we assume the average length of vertex names is some constant and hence, ignore it in time complexity analysis.} & $\blacktriangleright$ $\rhd$ {$\mathcal{O}(m^3 \cdot \log m)$} & $\mathcal{O}(m^2n)$\\
        % \rownumber &  &  & \\
        % \rownumber &  &  & \\
        % \rownumber &  &  & \\
        % \rownumber &  &  & \\
        % \rownumber &  &  & \\
        \hline
    \end{tabular}
    \caption{Line wise time complexity of Algorithm~\ref{alg:VC}. A highlight denotes the number of loop iterations. A pointer ($\blacktriangleright$) denotes that a line is within the loop. {Without loss of generality, we assume the average length of vertex names is a constant and hence, ignore it in time complexity analysis of Line 1.}}
    \label{tab:time_complexityAlg1}
\end{table}

\begin{table}
    \centering
    \begin{tabular}{|c||l|l|l|}
        \hline
        Line  Number& Line  complexity& Loop  complexity& Dominant complexity\\
        %Line  & Line  & Loop  & Dominant \\
        % Number &  complexity &  complexity &  complexity\\
        \hline
        \hline
        \rownumber & $\mathcal{O}(1)$ & - & $\mathcal{O}(1)$\\
        \rownumber & $\mathcal{O}(1)$ & \colorbox{blue!20}{$\mathcal{O}(m)$} & $\mathcal{O}(m)$\\
        \rownumber & $\mathcal{O}(1)$ & $\blacktriangleright$ \colorbox{blue!20}{$\mathcal{O}(m^2)$} & $\mathcal{O}(m^2)$\\
        \rownumber & $\mathcal{O}(n^2)$ & $\blacktriangleright$ $\rhd$ {$\mathcal{O}(m^2n^2)$} & $\mathcal{O}(m^2n^2)$\\
        \rownumber & $\mathcal{O}(1)$ & $\blacktriangleright$ $\rhd$ {$\mathcal{O}(m^2)$} & $\mathcal{O}(m^2n^2)$\\
        \rownumber & $\mathcal{O}(n^2)$ & $\blacktriangleright$ $\rhd$ {$\mathcal{O}(m^2n^2)$} & $\mathcal{O}(m^2n^2)$\\
        \rownumber & $\mathcal{O}(1)$ & $\blacktriangleright$ $\rhd$ {$\mathcal{O}(m^2)$} & $\mathcal{O}(m^2n^2)$\\
        \rownumber & $\mathcal{O}(n^2)$ & $\blacktriangleright$ $\rhd$ {$\mathcal{O}(m^2n^2)$} & $\mathcal{O}(m^2n^2)$\\
        \rownumber & $\mathcal{O}(1)$ & $\blacktriangleright$ $\rhd$ {$\mathcal{O}(m^2)$} & $\mathcal{O}(m^2n^2)$\\
        \rownumber & $\mathcal{O}(1)$ & $\blacktriangleright$ $\rhd$ {$\mathcal{O}(m^2)$} & $\mathcal{O}(m^2n^2)$\\
        \rownumber & $\mathcal{O}(n^2)$ & $\blacktriangleright$ $\rhd$ {$\mathcal{O}(m^2n^2)$} & $\mathcal{O}(m^2n^2)$\\
        \rownumber & - & - & $\mathcal{O}(m^2n^2)$\\
        \rownumber & $\mathcal{O}(m)$ & $\blacktriangleright$ $\rhd$ {$\mathcal{O}(m^3)$} & $\mathcal{O}(m^2n^2)$\\
        \rownumber & $\mathcal{O}(m+m^2)$ & $\blacktriangleright$ $\rhd$ {$\mathcal{O}(m^3+m^4)$} & $\mathcal{O}(m^2n^2)$\\
        \rownumber & $\mathcal{O}(n)$ & $\blacktriangleright$ $\rhd$ {$\mathcal{O}(m^2n)$} & $\mathcal{O}(m^2n^2)$\\
        \rownumber & $\mathcal{O}(m)$ & $\blacktriangleright$ $\rhd$ {$\mathcal{O}(m^3)$} & $\mathcal{O}(m^2n^2)$\\
        \rownumber & - & - & $\mathcal{O}(m^2n^2)$\\
        \rownumber & - & - & $\mathcal{O}(m^2n^2)$\\
        \rownumber & $\mathcal{O}(1)$ & - & $\mathcal{O}(m^2n^2)$\\
        \hline
    \end{tabular}
    \caption{Line wise time complexity of Algorithm~\ref{alg:MaM}. A highlight denotes the number of loop iterations. A pointer ($\blacktriangleright$) denotes that a line is within a loop. An additional pointer ($\rhd$) denotes a nested loop.}
    \label{tab:time_complexityAlg2}
\end{table}

\begin{table}
    \centering
    \begin{tabular}{|c||l|l|l|}
        \hline
        %Line  Number& Line  complexity& Loop  complexity& Dominant complexity\\
        Line  & Line  & Loop  & Dominant \\
        Number &  complexity &  complexity &  complexity\\
        \hline
        \hline
        \rownumber & $\mathcal{O}(1)$ & - & $\mathcal{O}(1)$\\
        \rownumber & $\mathcal{O}(m)$ & - & $\mathcal{O}(m)$\\
        \rownumber & $\mathcal{O}(1)$ & \colorbox{blue!20}{$\mathcal{O}(m)$} & $\mathcal{O}(m)$\\
        \rownumber & $\mathcal{O}(m^2)$ & $\blacktriangleright$ {$\mathcal{O}(m^3)$} & $\mathcal{O}(m^3)$\\
        \rownumber & - & - & $\mathcal{O}(m^3)$\\
        \rownumber & $\mathcal{O}(m^3)$ [Table~\ref{tab:time_complexityAlg4}] & $\blacktriangleright$ $\mathcal{O}(m^4)$ & $\mathcal{O}(m^4)$\\
        \rownumber & - & - & $\mathcal{O}(m^4)$\\
        \rownumber & - & - & $\mathcal{O}(m^4)$\\
        \rownumber & - & - & $\mathcal{O}(m^4)$\\
        \rownumber & $\mathcal{O}(1)$ & \colorbox{blue!20}{$\mathcal{O}(m)$} & $\mathcal{O}(m^4)$\\
        \rownumber & $\mathcal{O}(1)$ & $\blacktriangleright$ {$\mathcal{O}(m)$} & $\mathcal{O}(m^4)$\\
        \rownumber & $\mathcal{O}(1)$ & $\blacktriangleright$ {$\mathcal{O}(m)$} & $\mathcal{O}(m^4)$\\
        \rownumber & $\mathcal{O}(m)$ & $\blacktriangleright$ {$\mathcal{O}(m^2)$} & $\mathcal{O}(m^4)$\\
        \rownumber & $\mathcal{O}(m^2)$ & $\blacktriangleright$ {$\mathcal{O}(m^3)$} & $\mathcal{O}(m^4)$\\
        \rownumber & $\mathcal{O}(m^3)$ [Table~\ref{tab:time_complexityAlg4}] & $\blacktriangleright$ $\mathcal{O}(m^4)$ & $\mathcal{O}(m^4)$\\
        \rownumber & - & - & $\mathcal{O}(m^4)$\\
        \rownumber & - & - & $\mathcal{O}(m^4)$\\
        \rownumber & - & - & $\mathcal{O}(m^4)$\\
        \rownumber & - & - & $\mathcal{O}(m^4)$\\
        \rownumber & $\mathcal{O}(1)$ & \colorbox{blue!20}{$\mathcal{O}(m)$} & $\mathcal{O}(m^4)$\\
        \rownumber & $\mathcal{O}(1)$ & $\blacktriangleright$ {$\mathcal{O}(m)$} & $\mathcal{O}(m^4)$\\
        \rownumber & $\mathcal{O}(1)$ & $\blacktriangleright$ {$\mathcal{O}(m)$} & $\mathcal{O}(m^4)$\\
        \rownumber & $\mathcal{O}(1)$ & $\blacktriangleright$ {$\mathcal{O}(m)$} & $\mathcal{O}(m^4)$\\
        \rownumber & $\mathcal{O}(m^3)$ [Table~\ref{tab:time_complexityAlg4}] & $\blacktriangleright$ $\mathcal{O}(m^4)$ & $\mathcal{O}(m^4)$\\
        \rownumber & $\mathcal{O}(1)$ & $\blacktriangleright$ {$\mathcal{O}(m)$} & $\mathcal{O}(m^4)$\\
        \rownumber & - & - & $\mathcal{O}(m^4)$\\
        \rownumber & $\mathcal{O}(m^2)$ & $\blacktriangleright$ {$\mathcal{O}(m^3)$} & $\mathcal{O}(m^4)$\\
        \rownumber & $\mathcal{O}(m^3)$ [Table~\ref{tab:time_complexityAlg4}] & $\blacktriangleright$ $\mathcal{O}(m^4)$ & $\mathcal{O}(m^4)$\\
        \rownumber & $\mathcal{O}(m^2)$ & $\blacktriangleright$ {$\mathcal{O}(m^3)$} & $\mathcal{O}(m^4)$\\
        \rownumber & $\mathcal{O}(m^3)$ [Table~\ref{tab:time_complexityAlg4}] & $\blacktriangleright$ $\mathcal{O}(m^4)$ & $\mathcal{O}(m^4)$\\
        \rownumber & $\mathcal{O}(1)$ & $\blacktriangleright$ {$\mathcal{O}(m)$} & $\mathcal{O}(m^4)$\\
        \rownumber & $\mathcal{O}(m^3)$ [Table~\ref{tab:time_complexityAlg4}] & $\blacktriangleright$ $\mathcal{O}(m^4)$ & $\mathcal{O}(m^4)$\\
        \rownumber & - & - & $\mathcal{O}(m^4)$\\
        \rownumber & - & - & $\mathcal{O}(m^4)$\\
        \rownumber & - & - & $\mathcal{O}(m^4)$\\
        \rownumber & $\mathcal{O}(1)$ & - & $\mathcal{O}(m^4)$\\ 
        \hline
    \end{tabular}
    \caption{Line wise time complexity of Algorithm~\ref{alg:LM}.  A highlight denotes the number of loop iterations. A pointer ($\blacktriangleright$) denotes that a line is within a loop.}
    \label{tab:time_complexityAlg3}
\end{table}

\begin{table}
    \centering
    \begin{tabular}{|c||l|l|l|}
        \hline
        %Line  Number& Line  complexity& Loop  complexity& Dominant complexity\\
        Line  & Line  & Loop  & Dominant \\
        Number &  complexity &  complexity &  complexity\\
        \hline
        \hline
        \rownumber & $\mathcal{O}(m+m)$ & - & $\mathcal{O}(m)$\\
        \rownumber & $\mathcal{O}(m)$ & - & $\mathcal{O}(m)$\\
        \rownumber & $\mathcal{O}(1)$ & - & $\mathcal{O}(m)$\\
        \rownumber & $\mathcal{O}(m^2)$ & - & $\mathcal{O}(m^2)$\\
        \rownumber & $\mathcal{O}(m+m)$ & - & $\mathcal{O}(m^2)$\\
        \rownumber & $\mathcal{O}(m^2)$ & \colorbox{blue!20}{$\mathcal{O}(m)$} & $\mathcal{O}(m^2)$\\
        \rownumber & $\mathcal{O}(m^2)$ & $\blacktriangleright$$\mathcal{O}(m^3)$ & $\mathcal{O}(m^3)$\\
        \rownumber & - & - & $\mathcal{O}(m^3)$\\
        \rownumber & $\mathcal{O}(m^2)$ & \colorbox{blue!20}{$\mathcal{O}(m)$} & $\mathcal{O}(m^3)$\\
        \rownumber & $\mathcal{O}(m^2)$ & $\blacktriangleright$$\mathcal{O}(m^3)$ & $\mathcal{O}(m^3)$\\
        \rownumber & - & - & $\mathcal{O}(m^3)$\\
        \rownumber & $\mathcal{O}(1)$ & - & $\mathcal{O}(m^3)$\\
        \hline
    \end{tabular}
    \caption{Line wise time complexity of Algorithm~\ref{alg:FnR}. A highlight denotes the number of loop iterations. A pointer ($\blacktriangleright$) denotes that a line is within a loop.}
    \label{tab:time_complexityAlg4}
\end{table}

%\newpage
 %\input{tex_files/extras-deleteIt}

\section{Conclusion}
%Prior to this work, there was no known polynomial-time algorithm for an NP-complete problem. Here, we discover an algorithm that is an amalgamation of maximum matching, breadth-first search, maximal matching and local minimization for vertex cover problem on unweighted simple connected graphs. An interesting future avenue can be to work on an  algorithm and on the corresponding proof for vertex cover problem on unweighted undirected graphs. On the other hand, one can improve the time complexity of this algorithm, especially by improving the efficiency of Algorithm~\ref{alg:FnR}. Finally, i
We conjecture  that the VC problem can be solved efficiently. It implies that DiRe committees can be computed efficiently. Hence, achieving diversity and representation may be more \emph{efficient} than initially expected. %Also, indeed, P $=$ NP.

%indeed, a conclusion of this work is P $=$ NP.

% \subsection{Broader Impact}
% \label{sec:ethicalImplications}
%We do not expect any \emph{immediate} ethical implication of our work. Moreover, we do not foresee any \emph{immediate} positive or negative implications either. This is primarily because extrapolating our algorithm to hypergraphs itself seems non-trivial. Finally, transitioning to some complexity-independent cryptographic system (e.g., information-theoretically secure cryptography) may be ideal.

% \paragraph{Broader Impact:} 
% We do not expect major, \emph{immediate}, positive or negative, \emph{practical} implications of this work. It is primarily because extrapolating our algorithm to elections where candidates are divided into arbitrarily sized  arbitrary groups itself seems non-trivial (a.k.a. extrapolating our algorithm to hypergraphs itself seems non-trivial). 
\section*{Acknowledgement}
Blank for now.
%visualalgo -- thanks to them!!!

%Copernicus- helio instead of geo-centrism -- we like to project ourselves onto other things.

%separate science from the scientist!! - %https://www.vox.com/culture/2016/10/3/13147536/elena-ferrante-anita-raja-doxxing-controversy-explained

\bibliographystyle{alpha}  
\bibliography{references}

\begin{thebibliography}{Edm65}

\bibitem[Coo71]{cook1971complexity}
Stephen~A Cook.
\newblock The complexity of theorem-proving procedures.
\newblock In {\em Proceedings of the third annual ACM symposium on Theory of computing}, pages 151--158, 1971.

\bibitem[Edm65]{edmonds1965paths}
Jack Edmonds.
\newblock Paths, trees, and flowers.
\newblock {\em Canadian Journal of mathematics}, 17:449--467, 1965.

\bibitem[Kar72]{karp1972reducibility}
Richard~M Karp.
\newblock Reducibility among combinatorial problems.
\newblock In {\em Complexity of Computer Computations}, pages 85--103. Springer, 1972.

\bibitem[Kho19]{khot2019proof}
Subhash Khot.
\newblock On the proof of the 2-to-2 games conjecture.
\newblock {\em Current Developments in Mathematics}, 2019(1):43--94, 2019.

\bibitem[Lev73]{levin1973universal}
Leonid~Anatolevich Levin.
\newblock Universal sequential search problems.
\newblock {\em Problemy peredachi informatsii}, 9(3):115--116, 1973.

\end{thebibliography}

\newpage

\appendix
\section{DiRe Committee and Vertex Cover}
\label{sec:dcvcequi}

We formally show the equivalence between the DiRe Committee Feasibility problem and the Vertex Cover problem on unweighted, undirected graphs.

\begin{definition}[DiRe Committee Feasibility Problem (DiReCF)]\label{def:DiReCF}
We are given an instance of a committee election consisting of 
(i) a set of candidates $C$ who are divided into arbitrary groups $R \in \mathcal{R}$ each of size {at most} two and a diversity constraint $l_R$ that stipulates the selection of {at least} one candidate from each non-empty group ($l_R = 1$ for all $R \in \mathcal{R}$ where $|R|>0$, $l_R = 0$ otherwise) and 
(ii) a set of voters $O$ who are divided into arbitrary populations $P \in \mathcal{P}$ each approving {at most} two candidates $W_P$ and a representation constraint $l_P$ that stipulates the selection of {at least} one candidate from each population who has a non-empty set of approved candidates ($l_P = 1$ for all $P \in \mathcal{P}$ where $|W_P|>0$, $l_P = 0$ otherwise).

Given a committee size $k$ that is a non-negative integer, the goal of DiReCF is to determine whether there is a committee $W$ of size at most $k$ that satisfies the given constraints such that $|R \cap W| \geq l_R$ for all $R \in \mathcal{R}$ and $|W_P \cap W| \geq l_P$ for all $P \in \mathcal{P}$?
\end{definition}

To keep this section standalone, we again define the vertex cover problem:

\begin{definition}[Vertex Cover Problem (VC)]\label{def:MVCdecision2}
Given a graph $G = (V, E)$ consisting of a set of vertices $V$ and a collection $E$ of 2-element subsets of $V$ called edges, the vertex cover of the graph $G$ is a subset of vertices $S \subseteq V$ that includes at least one endpoint of every edge of the graph, i.e., for all $e \in E$, $e\cap S \neq \phi$.

Given a non-negative integer $k$, the goal of VC is to determine whether the graph $G$ has a vertex cover $S$ of size at most $k$?
\end{definition}

We now show that DiReCF and VC on unweighted, undirected graphs are equivalent by (i) reducing VC to DiReCF and (ii) reducing DiReCF to VC.

\begin{theorem}\label{thm:dcvcequi}
    DiReCF and VC are equivalent.
\end{theorem}

\begin{proof}
    We first give a polynomial-time reduction from VC to DiReCF.

    \paragraph{VC $\leq_P$ DiReCF:} We reduce an instance of vertex cover (VC) problem to an instance of DiReCF.
    We have one candidate $c_i \in C$ for each vertex $v_i \in V$. We have one candidate group $R \in \mathcal{R}$ consisting of two candidates $c_i$ and $c_j$ for each edge $e \in E$ that connects vertices $v_i$ and $v_j$. For each candidate group $R \in \mathcal{R}$, we set the diversity constraint $l_R$ to one. Additionally, for each edge $e \in E$ that connects vertices $v_i$ and $v_j$, we have a population of voters $P \in \mathcal{P}$ who approve of two candidates $c_i$ and $c_j$ in $W_P$. For each voter population $P \in \mathcal{P}$, we set the representation constraint $l_P$ to one. Finally, we set the target committee size to be $k$. 
    
    We have a vertex cover of size at most $k$ if and only if there is a committee of size at most $k$ that satisfies all the constraints. 
    
    ($\Rightarrow$) If an instance of the vertex cover problem is a yes instance, then the corresponding instance of DiReCF is a yes instance. This is because if there is a vertex cover $S$ of size $k$, then for each vertex $v_i \in S$, we have a candidate $c_i$ in committee $W$ who is in one or more candidate groups and is among the approved candidates for one or more populations. As each edge is covered by the vertex cover $S$, at least one candidate from each candidate group and from each voter populations' approved candidates is present in the committee $W$ of size $k$.
    
    ($\Leftarrow$) If there is a committee $W$ of size $k$ that satisfies all the constraints, then there is a vertex cover $S$ of size $k$. This is because for each $c_i \in W$, there is a vertex $v_i \in S$. Given that all constraints are satisfied by $W$, it implies all edges are covered by the vertex cover.

    %Next, we now give a polynomial-time reductions from DiReCF to VC.
    
    \paragraph{DiReCF $\leq_P$ VC:} We reduce an instance of DiReCF problem to an instance of the vertex cover (VC) problem. 
    We have one vertex $v_i \in V$ for each candidate $c_i \in C$. Next, we have an edge $e \in E$ for the following scenarios:
    \begin{itemize}
        \item for each candidate group $R \in \mathcal{R}$ that has candidates $c_i$ and $c_j$, we have an edge that connects $v_i$ and $v_j$.
        \item for each candidate group $R \in \mathcal{R}$ that has only one candidate $c_i$, we have an edge that connects $v_i$ with $v_i$. Basically, we have a loop.
        \item for each voter population $P \in \mathcal{P}$ that approves of candidates $c_i$ and $c_j$, we have an edge that connects $v_i$ and $v_j$.
         \item for each voter population $P \in \mathcal{P}$ that approves only one candidate $c_i$, we have an edge that connects $v_i$ with $v_i$. We again have a loop.
    \end{itemize}
    For the cases described above, we have the diversity constraint $l_R=1$ for all candidate groups $R \in \mathcal{R}$ where $|R|>0$. We have the representation constraint $l_P=1$ for all voter populations $P \in \mathcal{P}$ where $|W_P|>0$. The constraints correspond to the requirement that each edge must be covered ($e\cap S \neq \phi$).
    We do nothing for candidate groups of size zero and for voter populations who do not approve of any candidates. The corresponding constraints are set to zero and are henceforth ignored. Finally, we set the target committee size and the size of the vertex cover to $k$. 

    We have a committee of size at most $k$ that satisfies all the constraints if and only if there is a vertex cover of size at most $k$. 

    ($\Rightarrow$) If there is a committee $W$ of size $k$ that satisfies all the constraints, then for each candidate $c_i \in W$, there is a vertex $v_i$ in the vertex cover $S$ of size $k$. This is because we know that $|R \cap W| \geq l_R$ for all candidate groups $R \in \mathcal{R}$ and $|W_P \cap W| \geq l_P$ for all voter populations $P \in \mathcal{P}$. It implies that $|e\cap S|\geq 1$ for all edges $e \in E$, which means $e\cap S \neq \phi$.
    
    ($\Leftarrow$) If there is a vertex cover $S$ of size $k$, then there is a committee $W$ of size $k$ that satisfies all the constraints. Each edge covered by a vertex in $S$ implies each constraint being satisfied by a candidate in $W$.

    %Overall, as VC $\leq_P$ DiReCF and DiReCF $\leq_P$ VC, the two problems are equivalent.
\end{proof}

In summary, as VC $\leq_P$ DiReCF and DiReCF $\leq_P$ VC, the two problems are equivalent and can be used interchangeably. For technical simplicity, the paper uses VC instead of DiReCF.

\section{Related Work}
\label{sec:rw}

All NP-complete problems are ``equivalent'' from the perspective of computational complexity theory. Hence, any progress toward finding an efficient algorithm for any one NP-complete problem will have an impact on each and every NP-complete problem. However, there are thousands of known NP-complete problems and a literature review on each one of them is beyond the scope of this paper. Therefore, we focus our discussion on the literature review of the vertex cover problem. Specifically, we elaborate upon how our algorithm is fundamentally different from previous work on the vertex cover problem. 

\subsection{Approximation Algorithms and Restricted Graphs}
While there is an extremely rich line of work discussing the (i) hardness and hardness of approximation of the vertex cover problem, (ii) finding approximation algorithms\footnote{By ``algorithms'', we mean a polynomial-time (efficient) algorithm unless and until noted otherwise.} for restricted cases (e.g., graphs with bounded degree) and (iii) finding exact algorithms for restricted cases (e.g., bipartite graphs), there is no work initiated to find an exact algorithm for the vertex cover problem on graphs for which the problem is NP-complete\footnote{Approximation algorithms are actually for NP-hard problems. In this discussion, we use NP-hardness and NP-completeness interchangeably.}. Hence, to the best of our knowledge, there is no prior work relevant to our approach. Additionally, the paper builds upon the common fact that the endpoints of a maximal matching of a graph form a vertex cover.

\subsection{Parameterized Complexity}
Broadly speaking, parameterized complexity and in particular, fixed-parameter tractability is the study of the complexity of computational problems \emph{conditioned} on one or more parameters. In contrast, our algorithm is \emph{unconditional}. Moreover, our work does not build upon any known parameterized algorithms. 

\subsection{Blossom Algorithm}
We used the Blossom algorithm for our implementation. Hence, we cite it and not papers that improve upon the Blossom algorithm (e.g., a faster algorithm for maximum matching due to Micali and Vazirani\footnote{\url{https://ieeexplore.ieee.org/document/4567800} (last accessed: February 8, 2024)}). Moreover, the time complexity of the Blossom algorithm has no impact on the overall time complexity of the algorithm presented in our paper. Hence, implementing a faster algorithm for maximum matching is not needed.

\section{Vertex Cover on Unweighted Simple Connected Graphs}
\label{sec:vcuscnpc}

We now prove that the vertex cover (VC) problem on unweighted simple connected graphs is NP-complete.

\begin{definition}[Simple Graph]\label{def:simple}
A graph $G = (V, E)$ is said to be a simple graph if the graph (i) is undirected, (ii) has no loops, i.e., it has no edge that starts and ends at the same vertex and (iii) does not have more than one edge between any pair of vertices. 
\end{definition}

\begin{definition}[Connected Graph]\label{def:connected}
A graph $G = (V, E)$ is said to be a connected graph if, for each pair of vertices, there exists a path that connects the pair of vertices.
\end{definition}

\begin{theorem}\label{thm:vcuscnpc}
    The vertex cover (VC) problem on unweighted simple connected graphs is NP-complete.
\end{theorem}

\begin{proof}

    We first show the problem's membership in NP and then proceed to reduce from a known NP-hard problem. 
    
    \paragraph{Membership in NP:} The vertex cover (VC) problem on unweighted simple connected graphs is in NP. Given a candidate solution and an integer $k$, we can easily verify if the solution is a vertex cover of size at most $k$.

    \paragraph{NP-hardness:} We reduce from a known NP-hard problem, namely the vertex cover (VC) problem on unweighted undirected graphs. Specifically, we reduce an instance of the vertex cover problem on unweighted undirected graphs (VC1) to an instance of the vertex cover problem on unweighted simple connected graphs (VC2)\footnote{The terms VC1 and VC2 are used in this reduction only.}.

    For each vertex $v_i \in V$ in VC1, there is a vertex $v'_i \in V'$ in VC2. Next, for the edges, we have the following scenarios:
    \begin{itemize}
        \item there is an edge $e \in E$ in VC1 that connects two distinct vertices $v_i$ and $v_j$: there is a corresponding edge $e' \in E'$ in VC2 that connects two distinct vertices $v'_i$ and $v'_j$.
        \item there are multiple edges in VC1 that connects two distinct vertices $v_i$ and $v_j$: there is one edge $e' \in E'$ in VC2 that connects two distinct vertices $v'_i$ and $v'_j$.
        \item there is an edge in VC1 that loops over the same vertex $v_i$: create a dummy vertex $d'_i \in D'$ and then, there is an edge $e' \in E'$ in VC2 that connects the vertex $v'_i$ with the dummy vertex $d'_i$. Overall, for each loop in VC1, there is a dummy vertex created in VC2.        
     \end{itemize}
    
    Next, there is one dummy vertex $u' \in U'$ in VC2 that is connected to each vertex $v' \in V'$ and dummy vertex $d' \in D'$. Specifically, for each pair of vertices consisting of $u'$, there is a dummy edge $f' \in F'$ that connects the pair of vertices. In summary, the vertices in VC2 consist of a union of the following: $V' \cup D' \cup U'$. The edges in VC2 consist of a union of the following: $E' \cup F'$. Finally, we set the vertex cover size in VC2 to be at most $k+1$. 

\begin{figure}[t]
    \centering
    \includegraphics[width=1\linewidth]{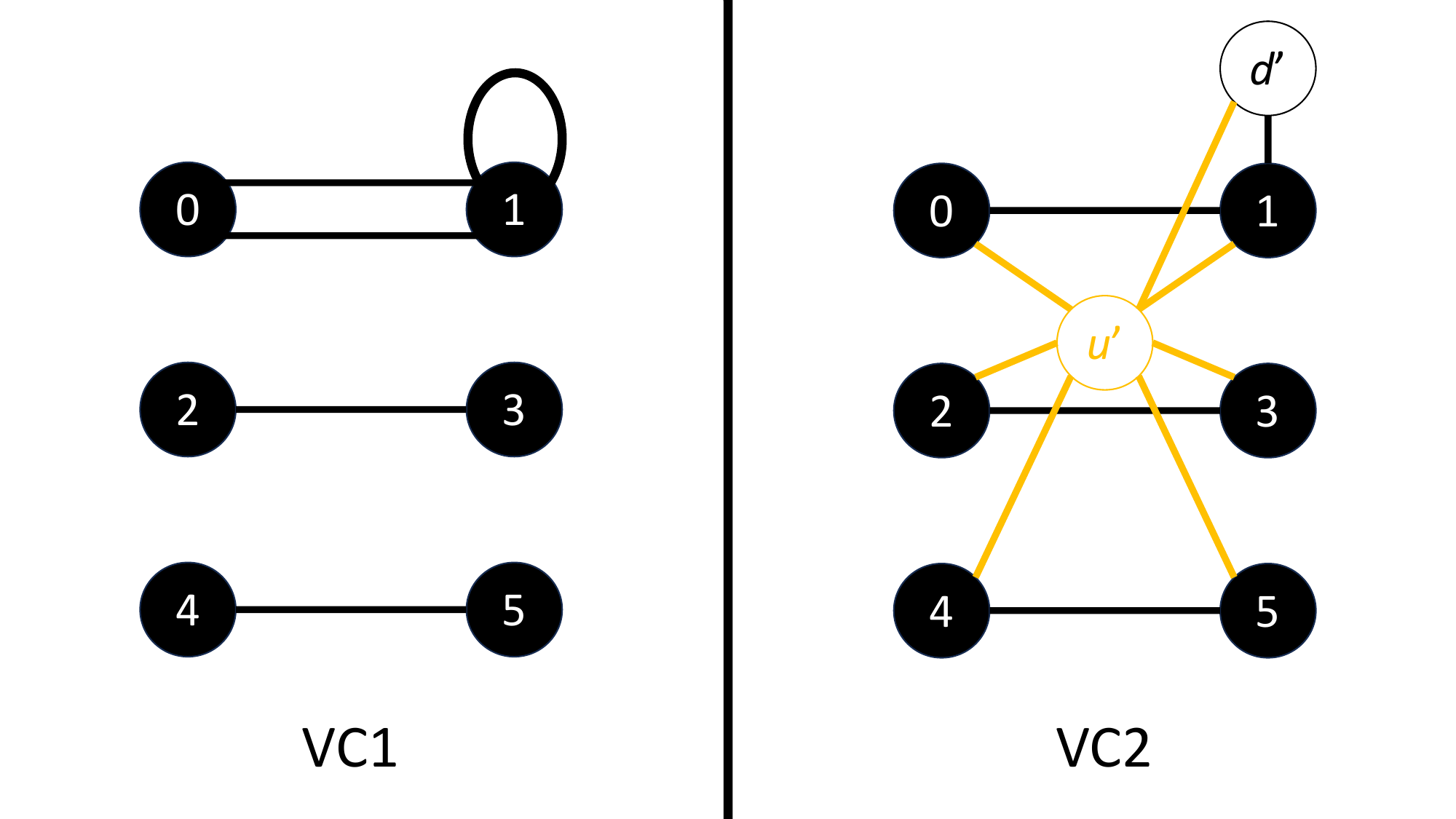}
    \caption{VC1 denotes an instance of the vertex cover problem on unweighted, undirected graph. VC2 denotes an instance of the vertex cover problem on unweighted simple connected graph. The multi-edges connecting vertices 0 and 1 in VC1 are removed in VC2. The loop connecting vertex 1 to itself is replaced by an edge in VC2 that connects vertex 1 to a dummy vertex $d'$. Another dummy vertex $u'$ (yellow vertex) is added to VC2 and is connected to all existing vertices to make the graph connected.}
    \label{fig:vcuscnpc}
\end{figure}
    
    It remains to be proven that there is a vertex cover on unweighted undirected graph of size at most $k$ if and only if there is a vertex cover on unweighted simple connected graph of size at most $k+1$. 

    ($\Rightarrow$)  If there is a vertex cover $S$ of size $k$ in an instance of VC1, then for each vertex $v_i \in S$, we have a vertex $v'_i$ in the vertex cover $S'$ of VC2. $S'$ covers all edges $e' \in E'$ of VC2. Additionally, dummy vertex $u'$ is always in the vertex cover $S'$, which covers all the edges $f' \in F'$. Consequently, the size of the vertex cover of VC2 is $k+1$.

\begin{table}
    \centering
    \begin{tabular}{|c|c|c||l|}
        \hline
        Dummy vertices $D'$& Vertices $V'$& Dummy vertex $U'$ & Case\\
        %Line  & Line  & Loop  & Dominant \\
        % Number &  complexity &  complexity &  complexity\\
        \hline
        \hline
        \xmark & \xmark & \xmark & Not possible\\
        \xmark & \cmark & \xmark & Case 3\\
        \cmark & \xmark & \xmark & Not possible\\
        \cmark & \cmark & \xmark & Not possible\\
        \xmark & \xmark & \cmark & Case 4\\
        \xmark & \cmark & \cmark & Case 1\\
        \cmark & \xmark & \cmark & Case 2\\
        \cmark & \cmark & \cmark & Case 2\\
       
        \hline
    \end{tabular}
    \caption{A summary of different possibilities of presence (\cmark) and absence (\xmark) of vertices from each set of vertices in the minimum vertex cover $S'$ in an instance of VC2. Each Case corresponds to an instance of VC2 being a yes instance in the proof of correctness in the reverse direction for Theorem~\ref{thm:vcuscnpc}.}
    \label{tab:proof_case_summary_vcuscnpc}
\end{table}
    
    ($\Leftarrow$)  The instance of the VC2 problem is a yes instance when each and every edge is covered. Then the corresponding instance of the VC1 problem is a yes instance as well. More specifically, there are the following cases when the instance of the VC2 problem can be a yes instance, i.e., it has a vertex cover $S'$ of size $k+1$:
    \begin{enumerate}
        \item vertex cover $S'$ consists of zero dummy vertex from $D'$, $k$ vertices from $V'$, one vertex $u'$ - This is a trivial case and the instance of the VC1 problem will have vertex cover $S$ consisting of vertex $v_i$ for every vertex $v'_i \in S'$. This will be of size $k$.
        \item vertex cover $S'$ consists of $x$\footnote{variable $x$ is an integer such that $1 \leq x \leq k$.} dummy vertex from $D'$, $k-x$ vertices from $V'$, one vertex $u'$ - For each dummy vertex $d' \in D'$ selected, the corresponding vertex $v'_i \in V'$ connected to the dummy vertex is not selected. Hence, given that the dummy vertex is of degree one, it can be swapped with the vertex it is connected to. This won't have any effect on the validity of the vertex cover $S'$. In summary, $x$ dummy vertices from $D'$ in vertex cover $S'$ are replaced by the corresponding $x$ vertices from $V'$. Consequently, an instance of the VC1 problem will have vertex cover $S$ consisting of vertex $v_i$ for every vertex $v'_i \in S'$. This will be of size $k$.
        \item vertex cover $S'$ consists of zero dummy vertex from $D'$, $k+1$ vertices from $V'$ (vertex $u'$ is not selected) - This case may arise when VC2 is a complete graph. Specifically, when VC2 is a complete graph, it does not consist of any vertex in $D'$. Moreover, the vertex cover $S'$ of VC2 is equivalent to the vertex set $V'$. Hence, we can replace any vertex from $S'$ with dummy vertex $u'$ and the instance of VC2 still remains a yes instance. Formally, the new vertex cover $S''$ will consist of $\{S' \setminus \{v'\}\} \cup \{u'\}$ for some $v' \in V'$. Hence, for every $v' \in S''$ where $v' \in V'$, there is a corresponding $v \in S$ in VC1. The vertex cover $S$ in VC1 is of size $k$ as the new vertex cover $S''$ consists of $k$ vertices from $V'$.
        % \begin{itemize}
        %     \item VC2 is a complete graph: When VC2 is a complete graph, it does not consist of any vertex in $D'$. Moreover, the vertex cover $S'$ of VC2 is equivalent to the vertex set $V'$. Hence, we can replace any vertex from $S'$ with dummy vertex $u'$ and the instance of VC2 still remains a yes instance. Formally, the new vertex cover $S''$ will consist of \{$S' \setminus v' \cup u'$\} for some $v' \in V'$. Hence, for every $v' \in S''$ where $v' \in V'$, there is a corresponding $v \in S$ in VC1. The vertex cover $S$ in VC1 is of size $k$ as the new vertex cover $S''$ consists of $k$ vertices from $V'$.
        %     \item All vertices within the set $V' \cup U'$ of VC2 are connected to each other: When all vertices within the set $V' \cup U'$ of VC2 are connected to each other, the dummy vertices in $D'$ can not 
        % \end{itemize}
        \item vertex cover $S'$ consists of zero dummy vertex from $D'$, zero vertices from $V'$ and one vertex $u'$ - In such a case, one endpoint of all edges in VC2 is $u'$. Hence, the corresponding instance of VC1 contains no edges and its vertex cover will be a null set. 
    \end{enumerate} 
    Finally, note that no other cases can lead to a yes instance of VC2 (e.g., $S'$ is not a vertex cover if $S'$ consists of, for example, $x$ dummy vertices from $D'$, $k-x+1$ vertices from $V'$ and zero vertex from $U'$). 

    This completes the other direction of the proof of correctness. In turn, this completes the entire proof.
\end{proof}
\section{Implementation of Algorithm}
\label{sec:example}

We give an example to explain the implementation of the entire algorithm. Additional examples can be found \href{https://docs.google.com/presentation/d/1FJscULC9PZA4am6KJJgP6V1v2YPTrrPB/edit?usp=sharing&ouid=116854247679466528662&rtpof=true&sd=true}{\textcolor{blue}{\ul{here}}} and \href{https://docs.google.com/presentation/d/10bKCa4lX0fwvEVpaDMJ_YGmcVDFO_kqb/edit?usp=sharing&ouid=116854247679466528662&rtpof=true&sd=true}{\textcolor{blue}{\ul{here}}} (link will open to Google Slides).

\begin{figure}[h]
    \centering
    \includegraphics[width=0.96925\linewidth]{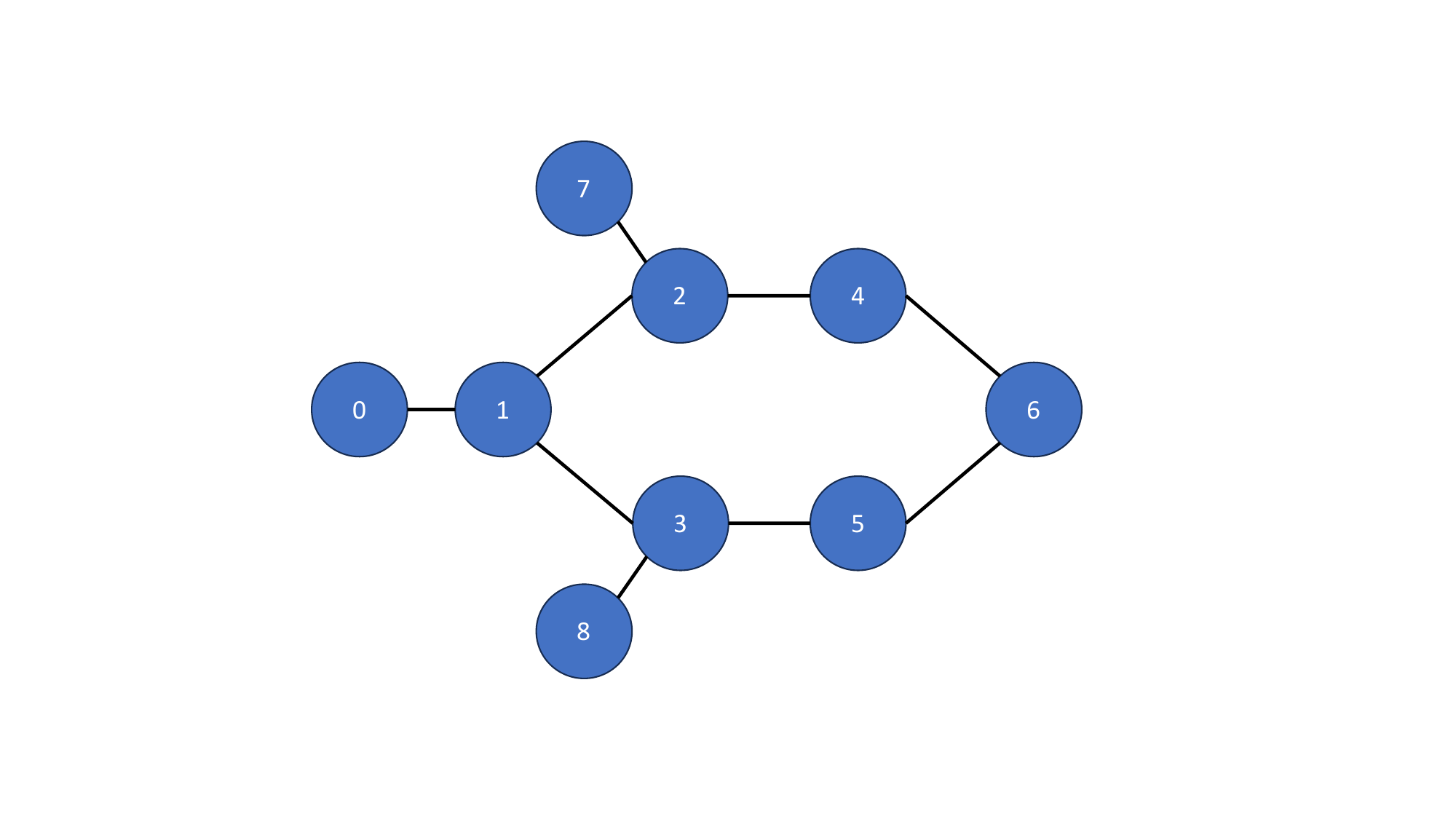}
    \caption{Example Graph $G$.}
    \label{fig:enter-label1}
\end{figure}

\begin{figure}[h]
    \centering
    \includegraphics[width=0.96925\linewidth,trim={0cm 0cm 0cm 3.25cm},clip]{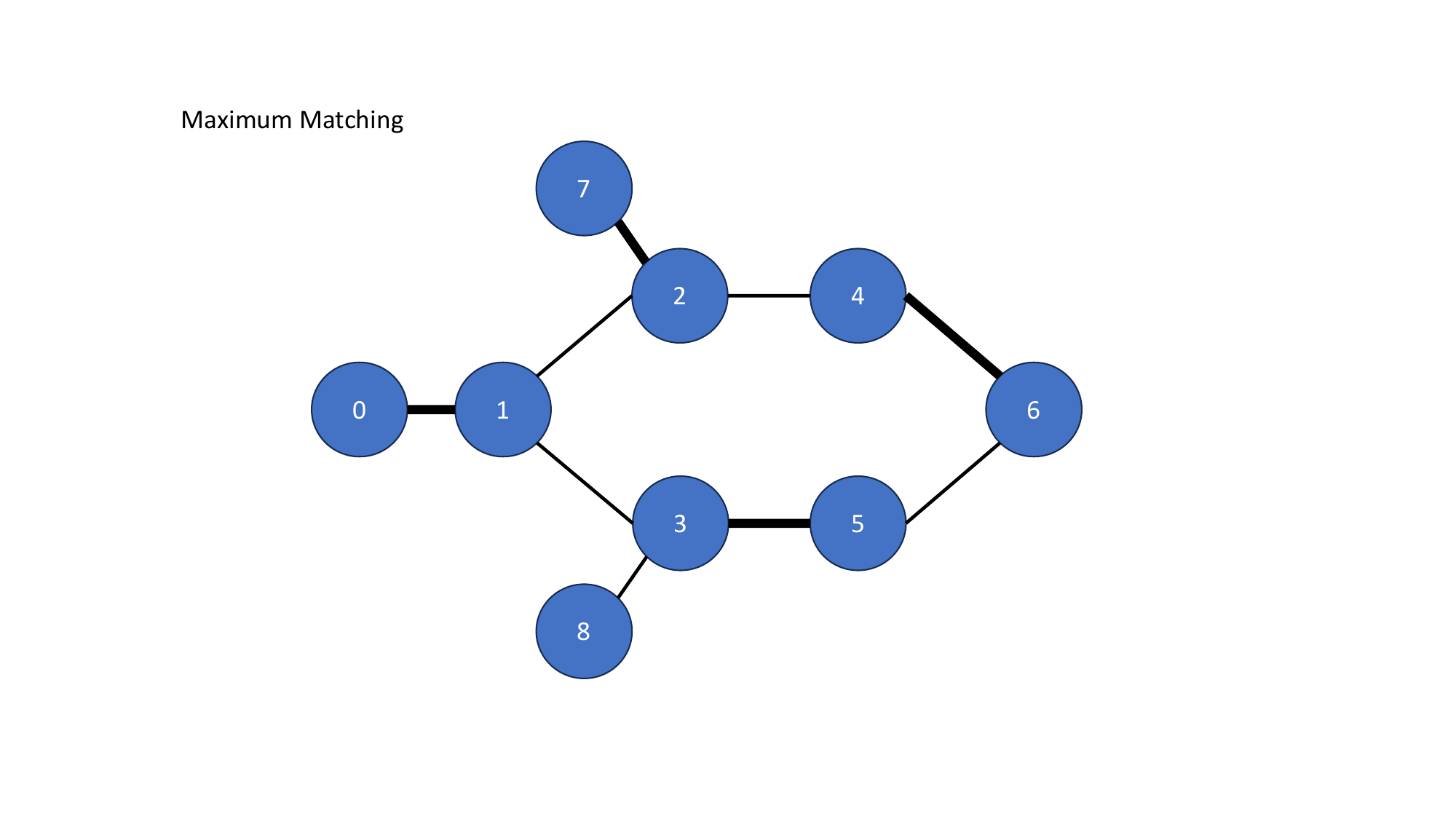}
    \caption{Bold edges \{(0, 1), (2, 7), (3, 5), (4, 6)\} form a maximum matching of graph $G$.}
    \label{fig:enter-label2}
\end{figure}

\begin{figure}
    \centering
    \includegraphics[width=1.1\linewidth]{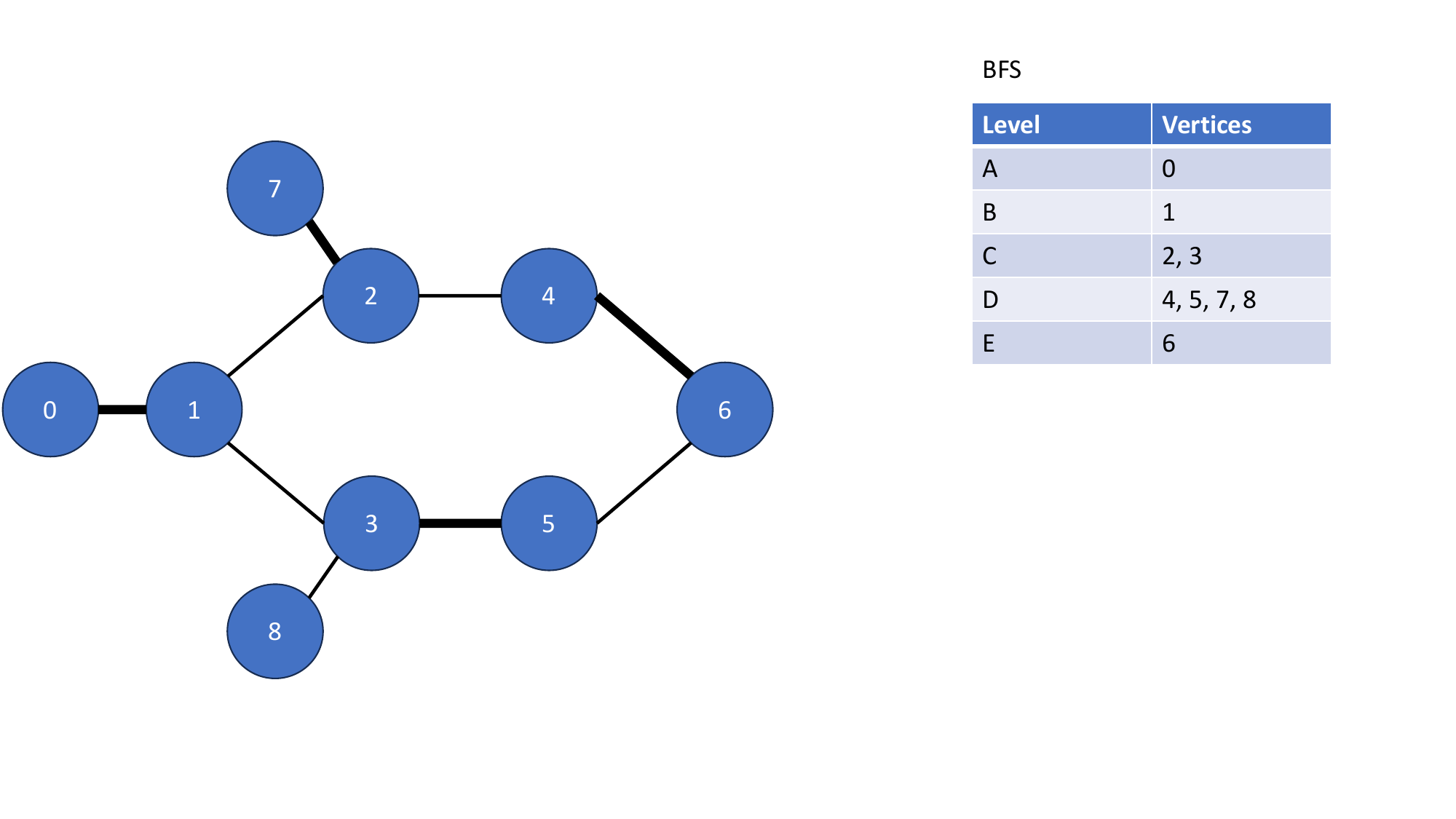}
    \caption{The ``BFS'' table lists the vertices at each level of the BFS (seeded on vertex `0').}
    \label{fig:enter-label3}
\end{figure}

\begin{figure}
    \centering
    \includegraphics[width=1.1\linewidth]{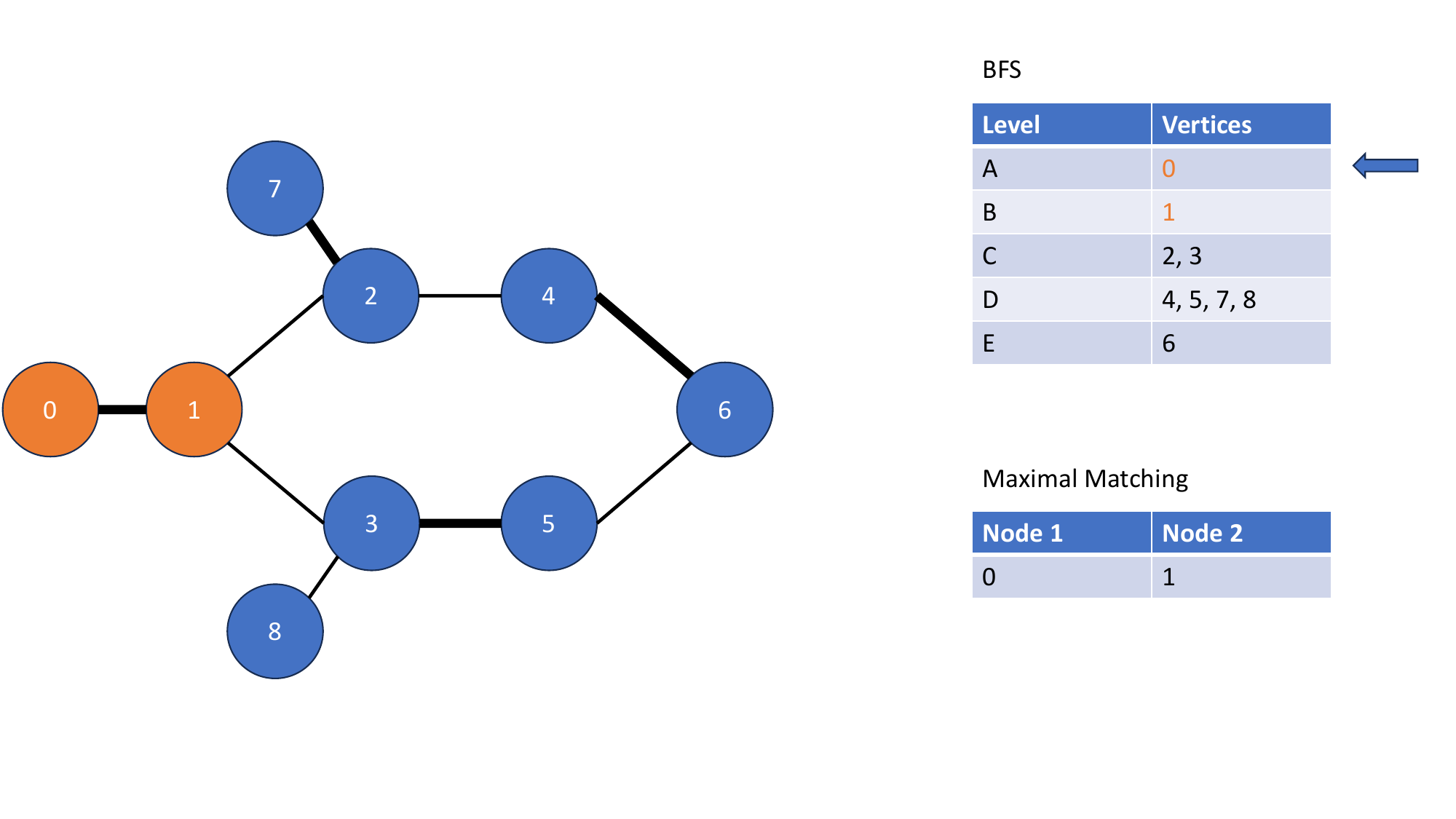}
    \caption{``Maximal Matching'' table lists vertices 0 and 1 (orange vertices in graph $G$), which are the endpoints of the first edge selected during maximal matching. Each endpoint is marked as visited (orange font; BFS table).}
    \label{fig:enter-label4}
\end{figure}

\begin{figure}
    \centering
    \includegraphics[width=1.1\linewidth]{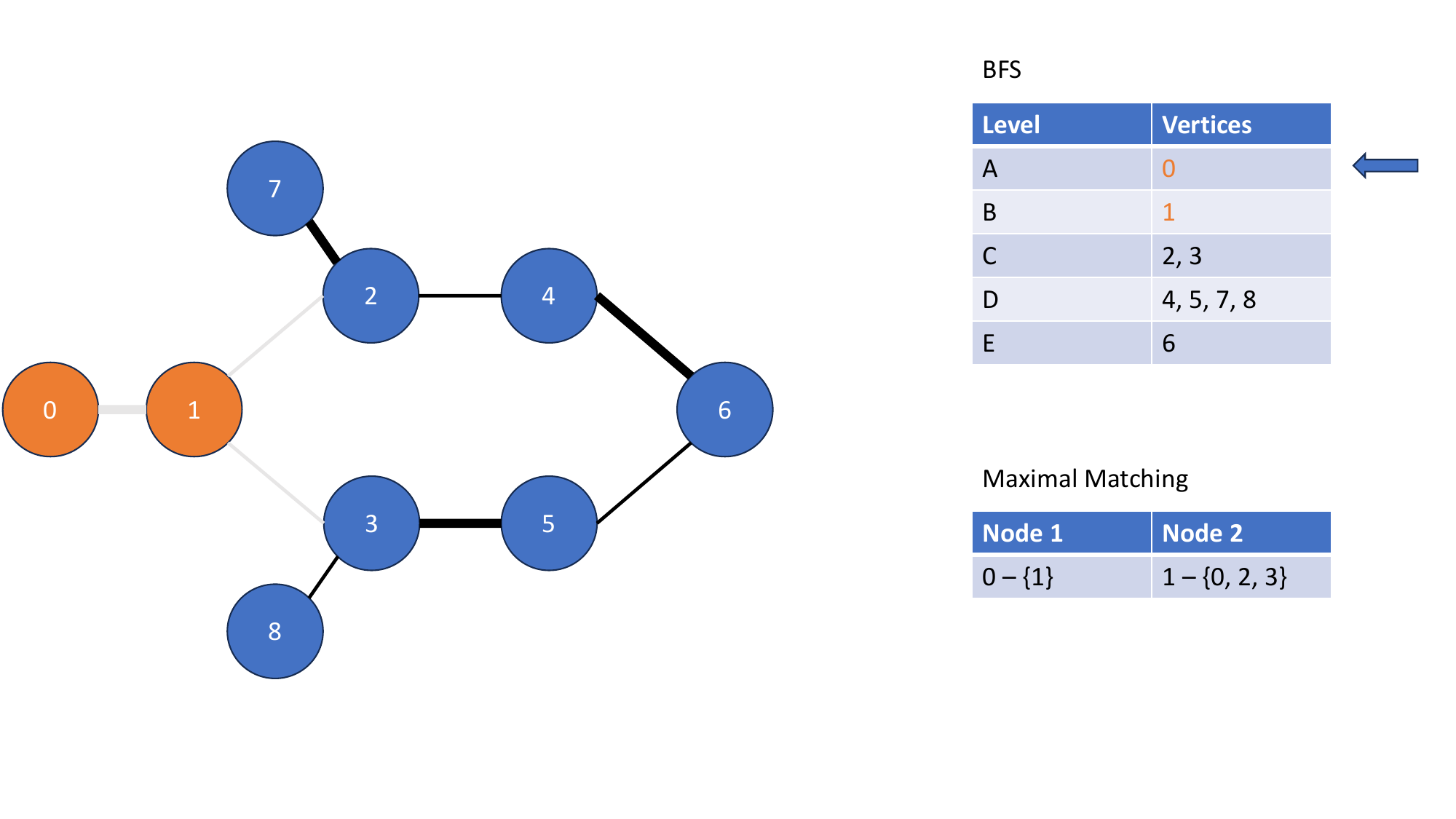}
    \caption{For each of the endpoints, namely 0 and 1, the respective curly brackets (\{\}) enlists the vertices connected to the corresponding vertex. Here, 0 is connected to \{1\} and 1 is connected to \{0, 2, 3\}. In graph $G$, the grayed out edges represent the removed edges.}
    \label{fig:enter-label5}
\end{figure}

\begin{figure}
    \centering
    \includegraphics[width=1.1\linewidth]{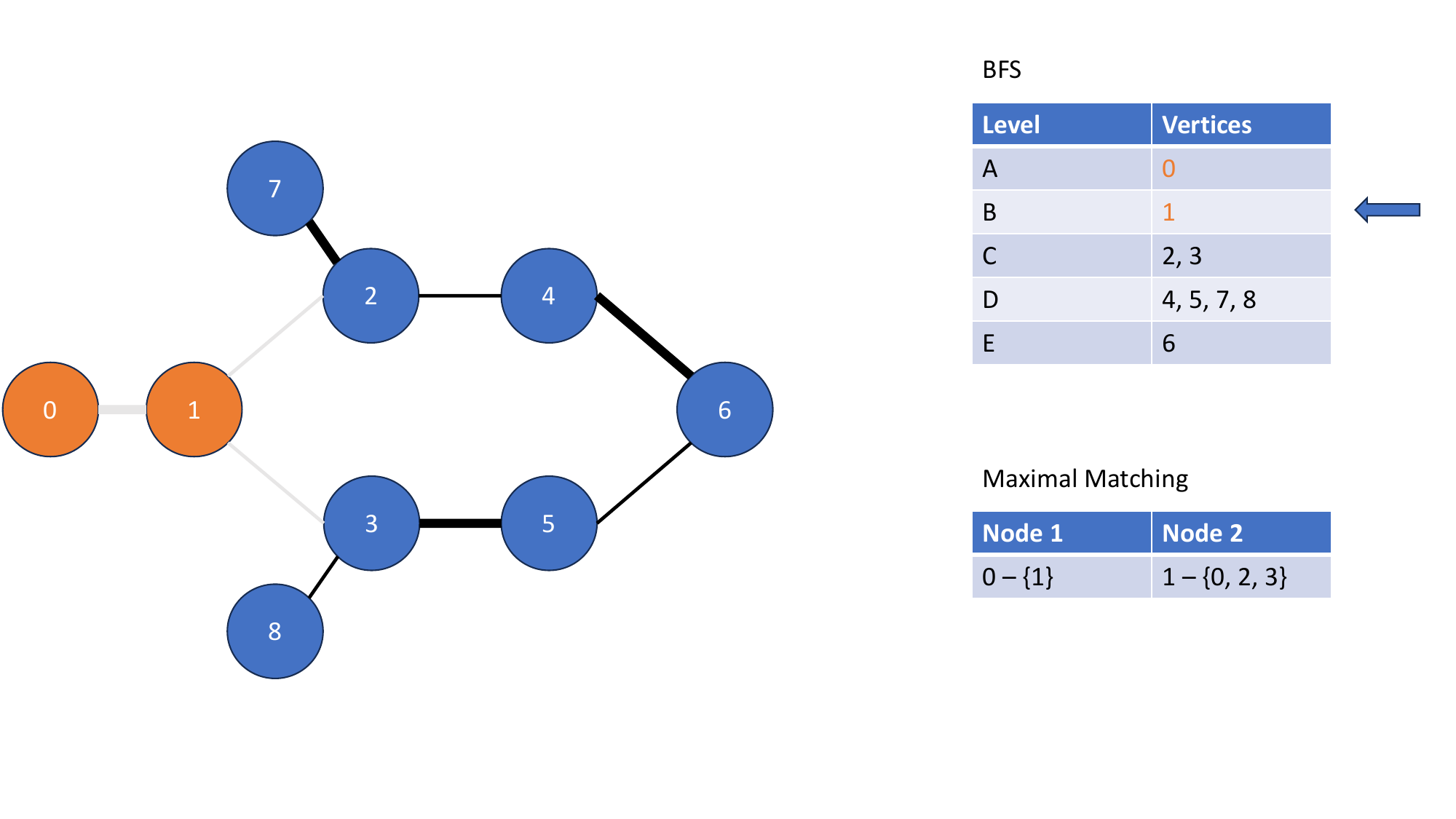}
    \caption{As all vertices on Level A of  BFS table is visited, the pointer now is on Level B.}
    \label{fig:enter-label6}
\end{figure}

\begin{figure}
    \centering
    \includegraphics[width=1.0875\linewidth]{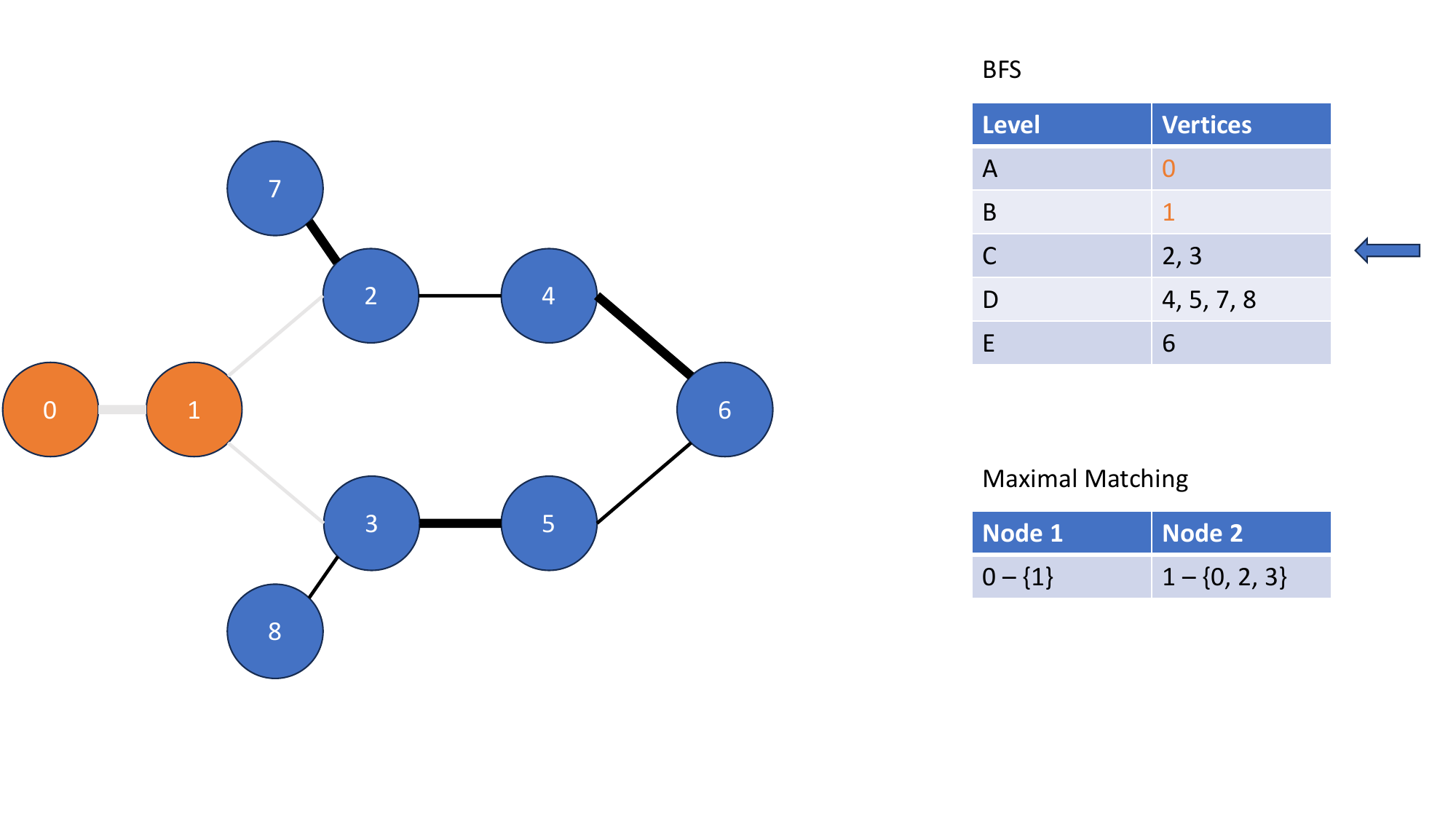}
    \caption{As all vertices on Level B of BFS table is visited, the pointer now is on Level C.}
    \label{fig:enter-label7}
\end{figure}

\begin{figure}
    \centering
    \includegraphics[width=1.0875\linewidth]{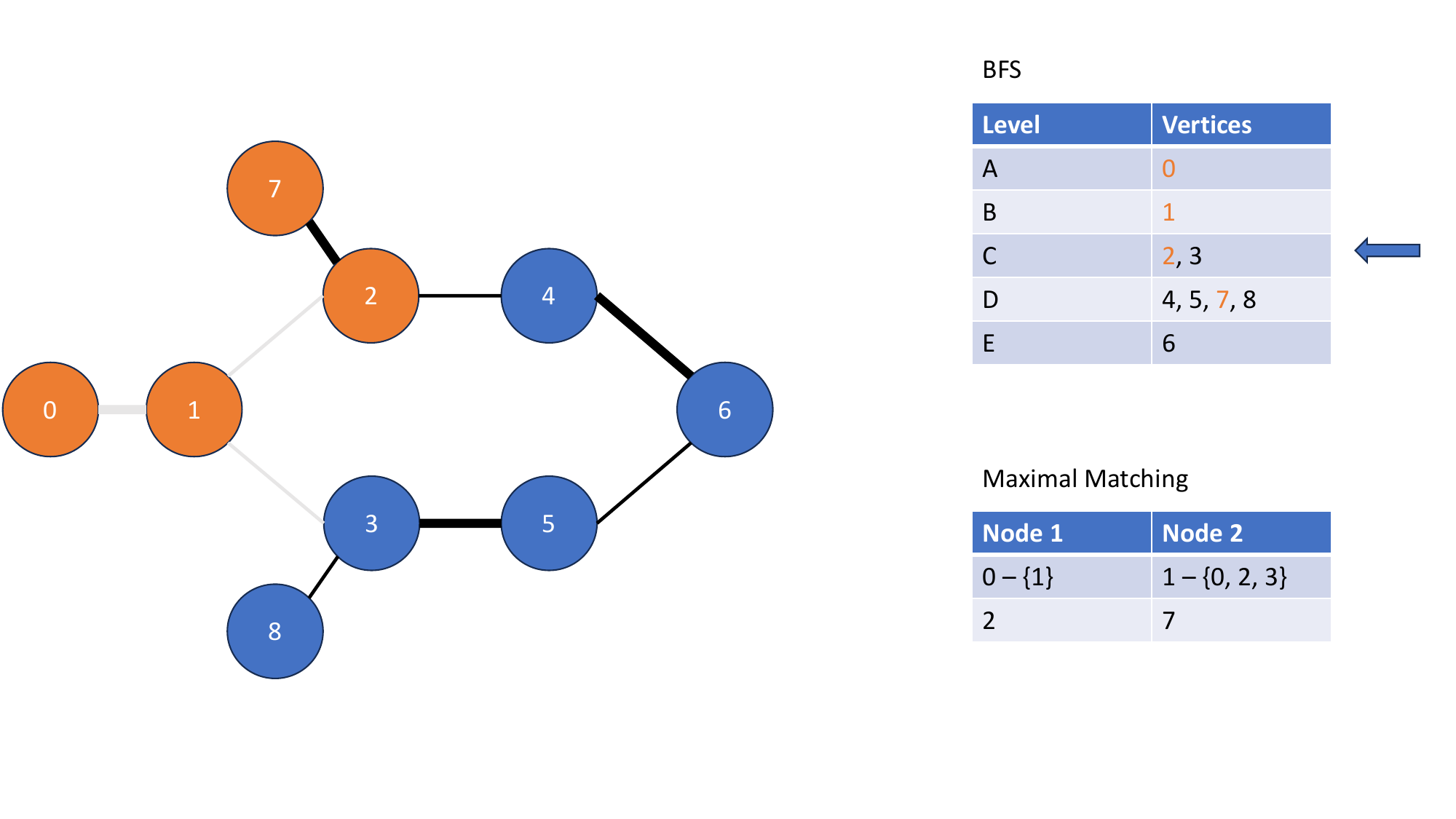}
    \caption{Vertex 2 comes before vertex 3 when sorted lexicographically. Hence, it is selected as one of the endpoints. As the edge connecting vertices 2 and 7 is part of maximum matching, it is preferred over edge connecting vertices 2 and 4. Hence, ``Maximal Matching'' table lists vertices 2 and 7, which are the endpoints of the second edge selected during maximal matching. Each endpoint is marked as visited (orange font; BFS table).}
    \label{fig:enter-label8}
\end{figure}

\begin{figure}
    \centering
    \includegraphics[width=1.05\linewidth]{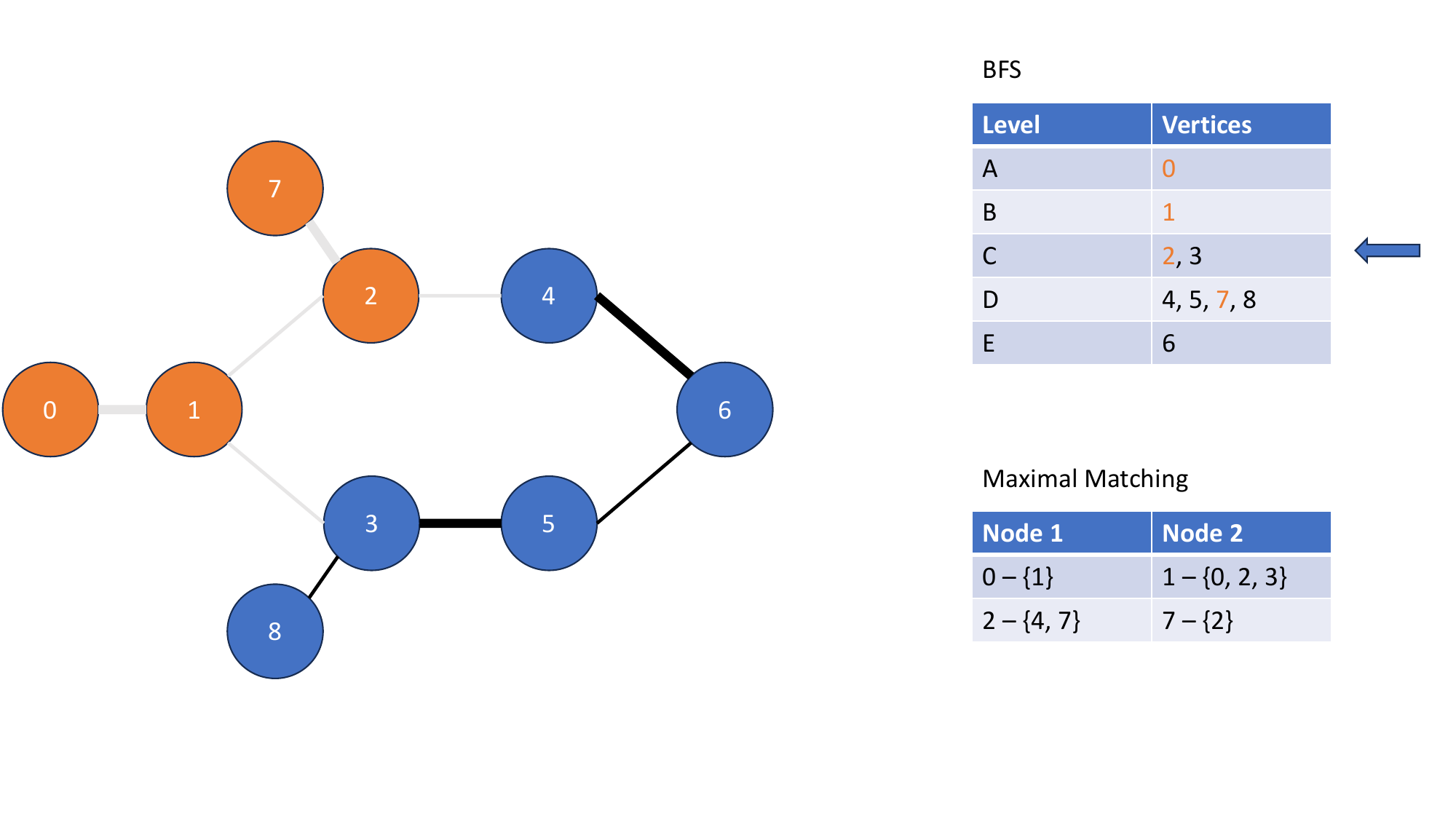}
    \caption{For each of the endpoints, namely 2 and 7, the respective curly brackets (\{\}) enlists the vertices connected to the corresponding vertex via an unremoved edge. Here, 2 is connected to \{4, 7\} and 7 is connected to \{2\}. The corresponding edges are removed (grayed out).}
    \label{fig:enter-labe9}
\end{figure}

\begin{figure}
    \centering
    \includegraphics[width=1.05\linewidth]{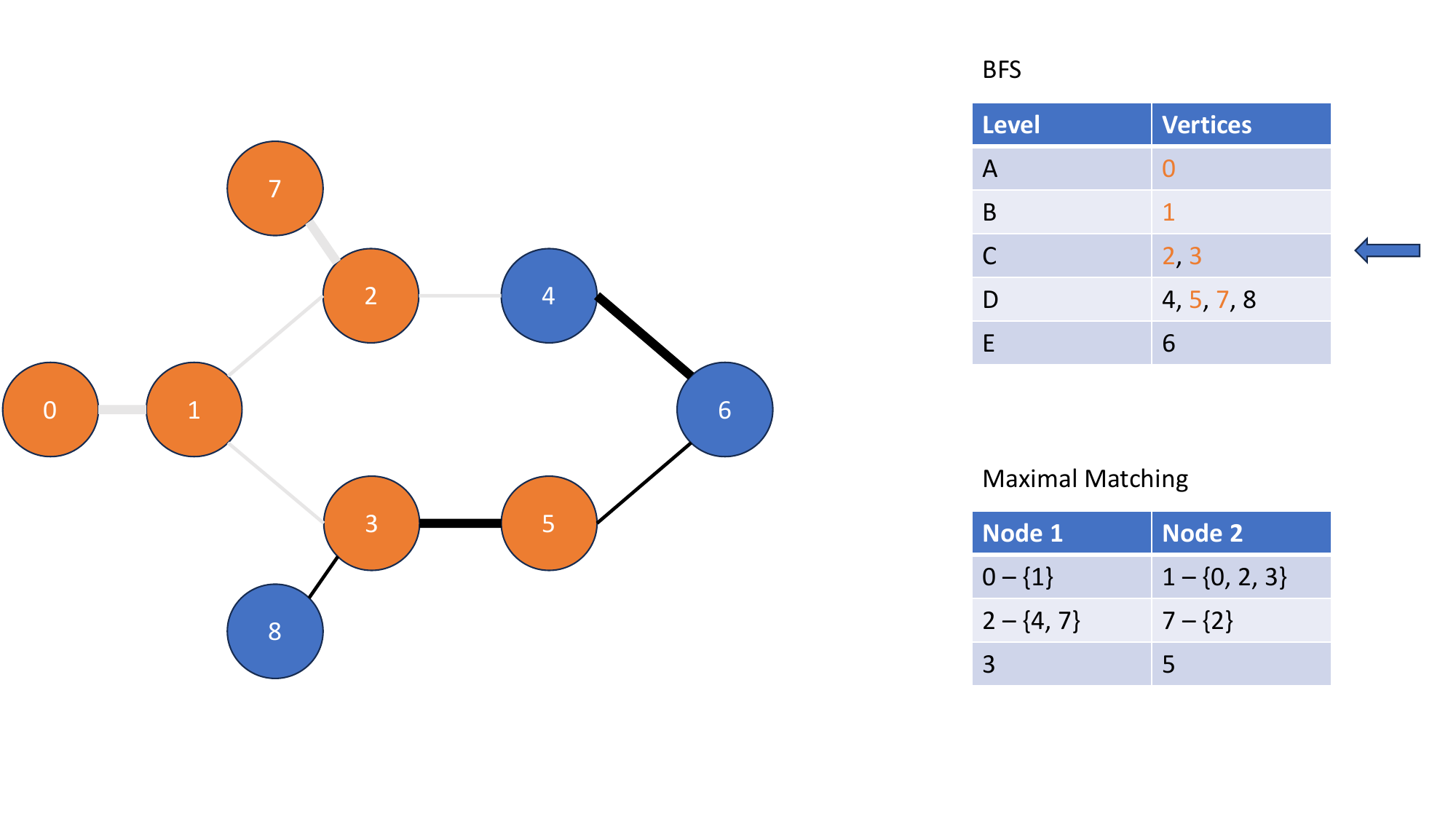}
    \caption{As the edge connecting vertices 3 and 5 is part of maximum matching, it is preferred over edge connecting vertices 3 and 8. Hence, ``Maximal Matching'' table lists vertices 3 and 5, which are the endpoints of the third edge selected during maximal matching. Each endpoint is marked as visited (orange font; BFS table).}
    \label{fig:enter-label10}
\end{figure}

\begin{figure}
    \centering
    \includegraphics[width=1.1\linewidth]{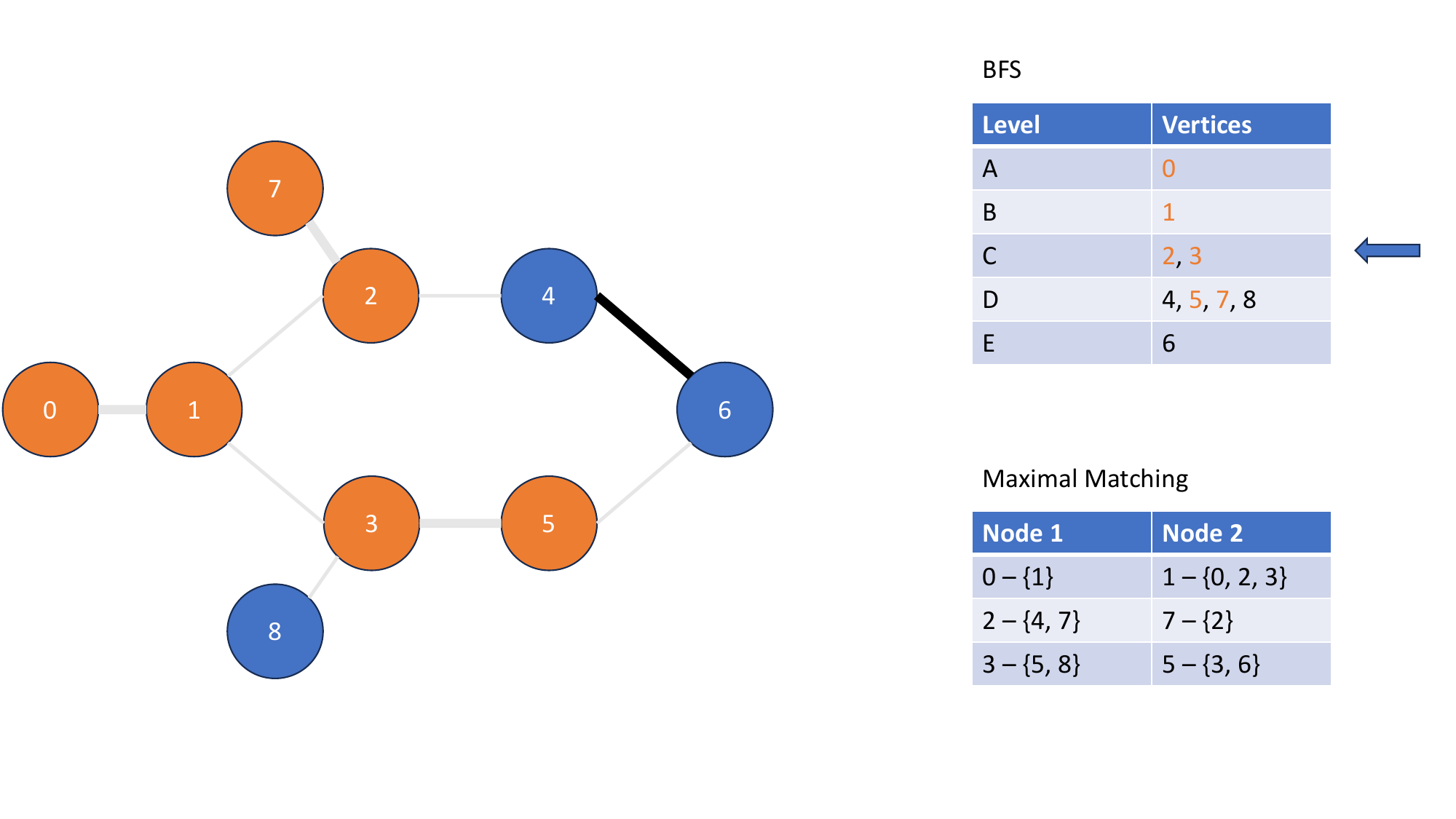}
    \caption{For each of the endpoints, namely 3 and 5, the respective curly brackets (\{\}) enlists the vertices connected to the corresponding vertex via an unremoved edge. Here, 3 is connected to \{5, 8\} and 5 is connected to \{3, 6\}. The corresponding edges are removed (grayed out).}
    \label{fig:enter-label11}
\end{figure}

\begin{figure}
    \centering
    \includegraphics[width=1.1\linewidth]{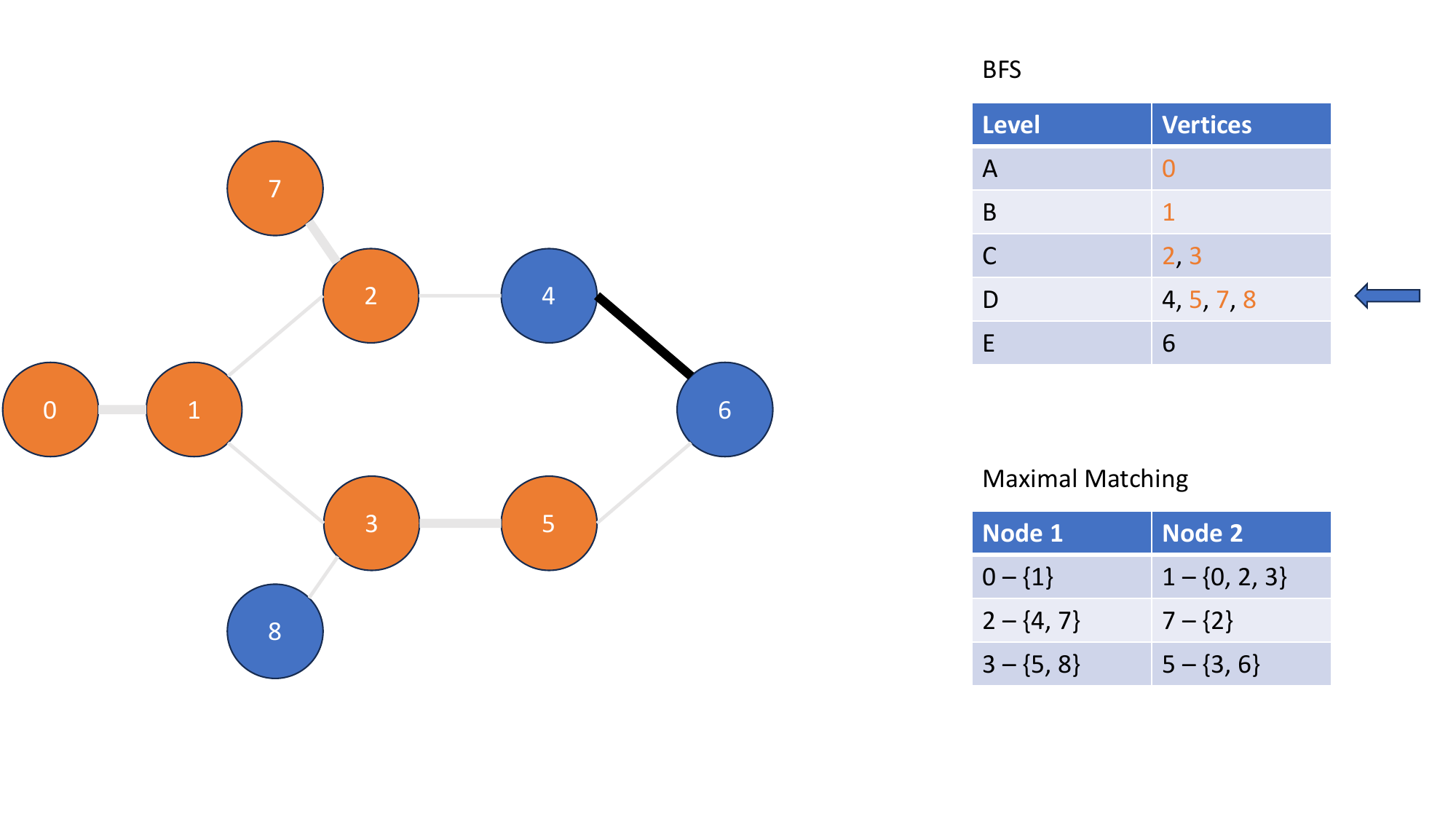}
    \caption{Vertex 8, which now has no unremoved edges, is marked as visited (orange font; BFS table). As all vertices on Level C of BFS table is visited, the pointer now is
on Level D.}
    \label{fig:enter-label12}
\end{figure}

\begin{figure}
    \centering
    \includegraphics[width=1.05\linewidth]{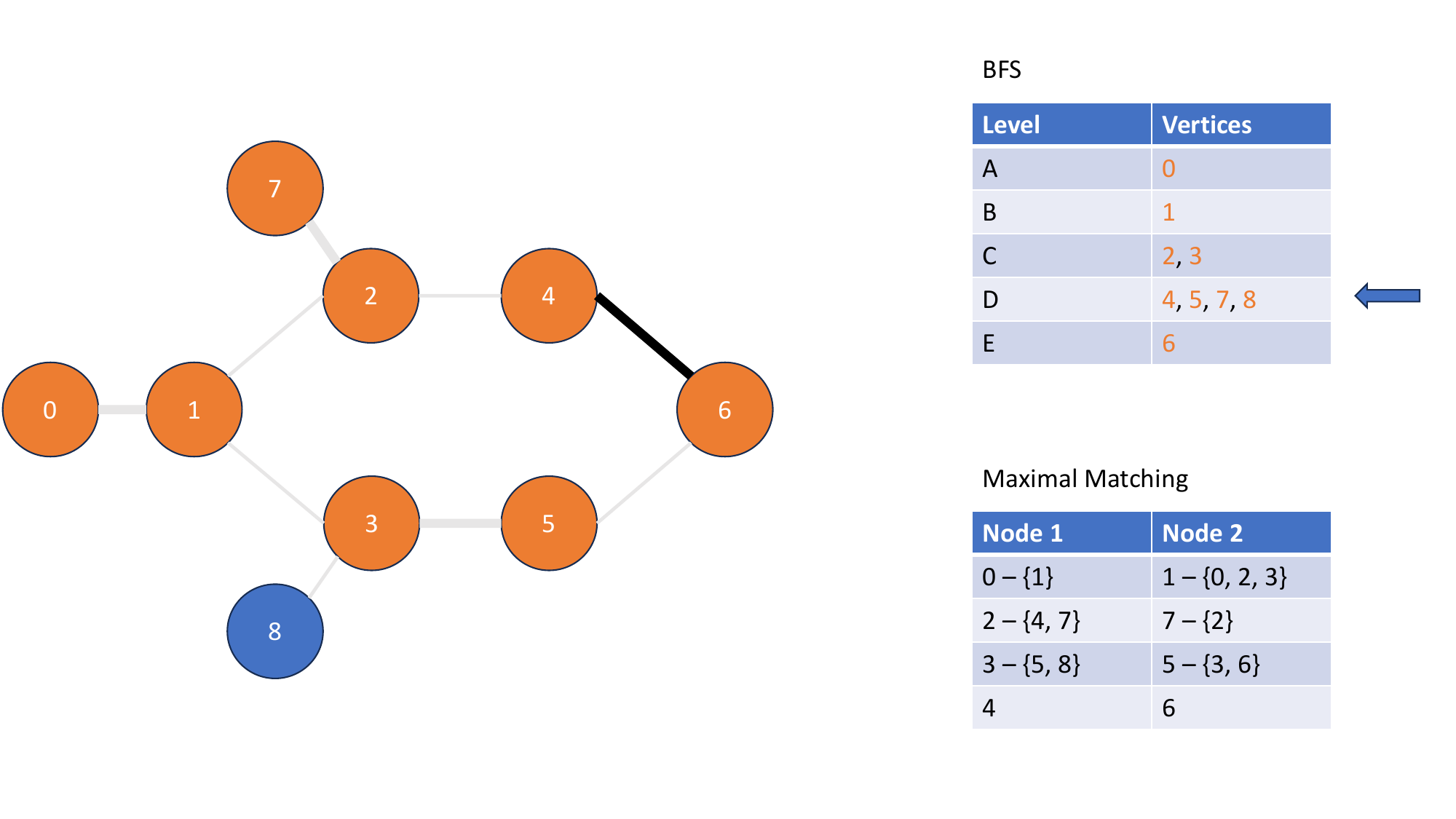}
    \caption{The edge connecting vertices 4 and 6, which is part of maximum matching, is the only remaining edge. Hence, ``Maximal Matching'' table lists vertices 4 and 6, which are the endpoints of the fourth and final edge selected during maximal matching. Each endpoint is marked as visited (orange font; BFS table).}
    \label{fig:enter-label13}
\end{figure}

\begin{figure}
    \centering
    \includegraphics[width=1.05\linewidth]{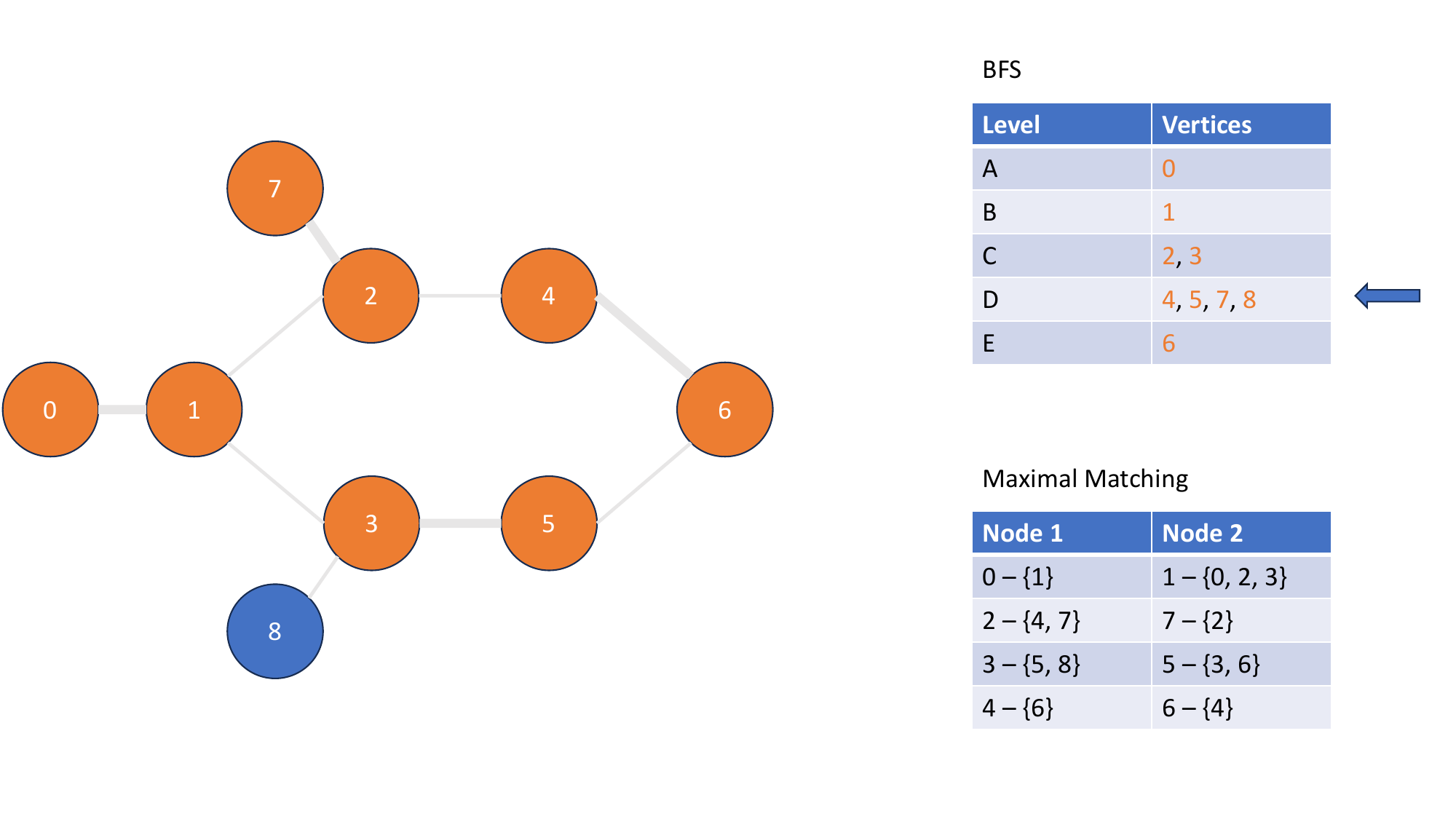}
    \caption{For each of the endpoints, namely 4 and 6, the respective curly brackets (\{\}) enlists the vertices connected to the corresponding vertex via an unremoved edge. Here, 4 is connected to \{6\} and 6 is connected to \{4\}. The corresponding edge is removed (grayed out).}
    \label{fig:enter-label14}
\end{figure}

\begin{figure}
    \centering
    \includegraphics[width=1.15\linewidth,trim={7cm 5cm 2cm 5cm},clip]{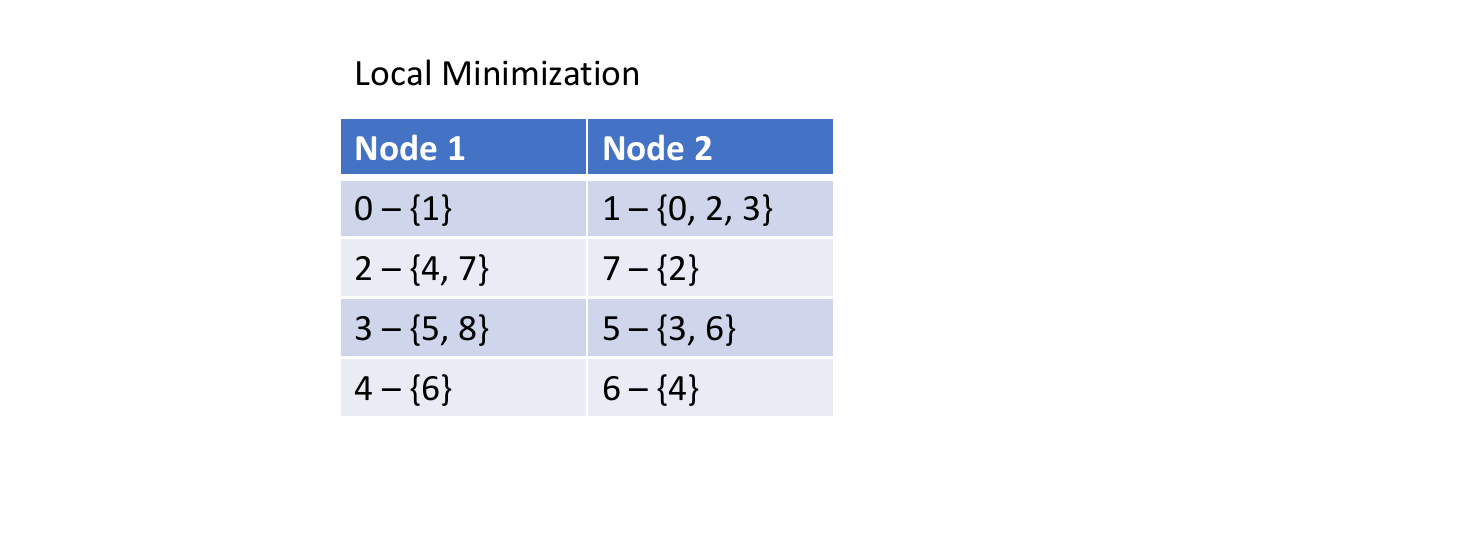}
    \caption{The ``Maximal Matching'' table will be used for ``Local Minimization'' phase of the algorithm. Vertices \{0,  1,  2, 3,  4, 5, 6, 7\} are labeled as eponymous ``endpoint'' vertices.}
    \label{fig:enter-label15}
\end{figure}

\begin{figure}
    \centering
    \includegraphics[width=1.15\linewidth,trim={7cm 5cm 2cm 5cm},clip]{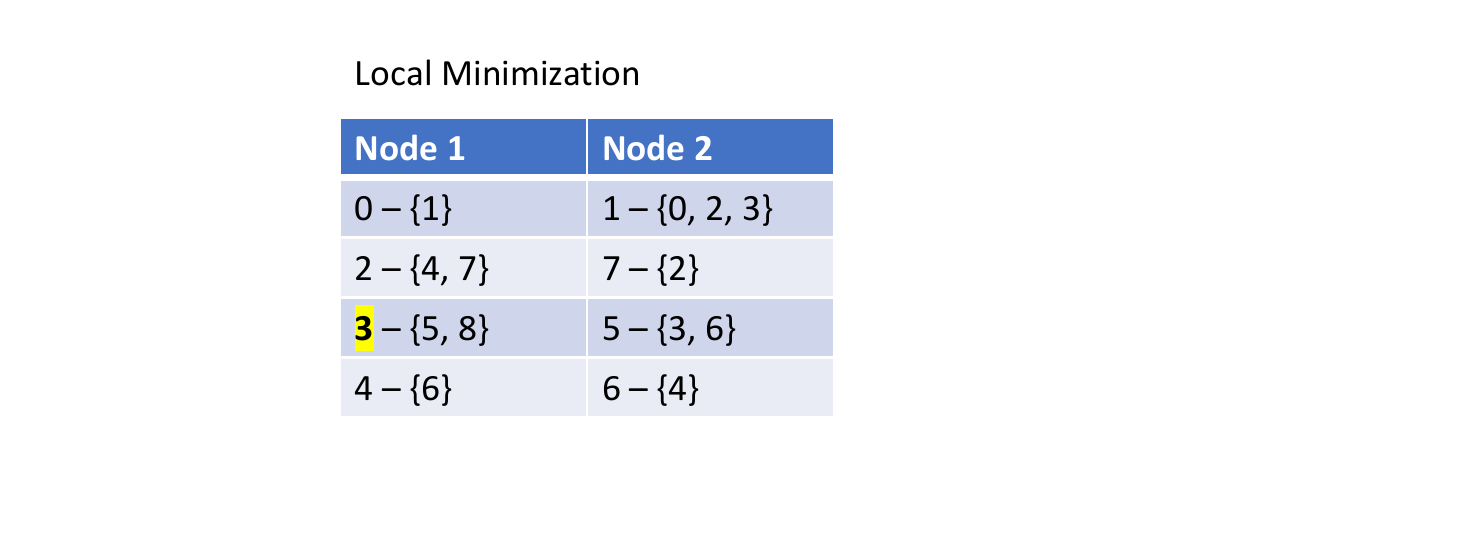}
    \caption{Vertex 3 is frozen (highlighted yellow) as it represents vertex 8, which is not an ``endpoint'' vertex. By default, the represents list of a frozen vertex (here, vertex 3) is removed (not shown here for convenience). }
    \label{fig:enter-label16}
\end{figure}

\begin{figure}
    \centering
    \includegraphics[width=1.15\linewidth,trim={7cm 5cm 2cm 5cm},clip]{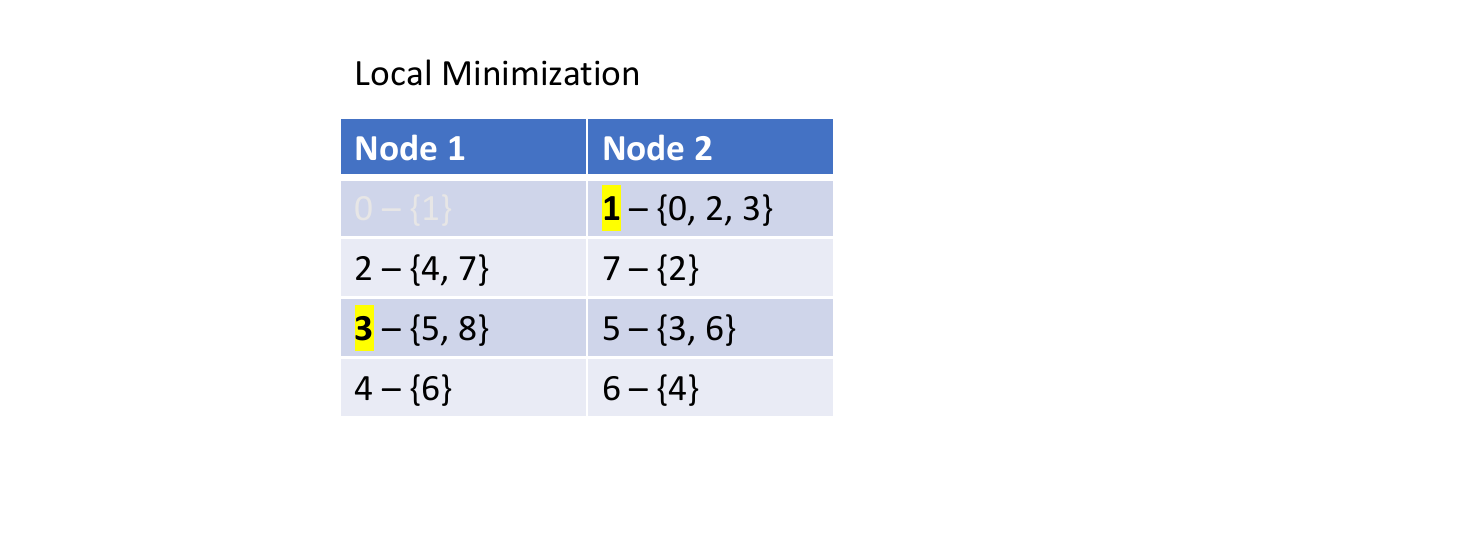}
    \caption{Vertex 0 in row 1 is removed (grayed out) as it  represents no other vertex except its same-row neighbor (namely vertex 1). Consequently, vertex 1 in row 1 is frozen (highlighted yellow) as it now represents a removed vertex  (namely vertex 0).}
    \label{fig:enter-label17}
\end{figure}

\begin{figure}
    \centering
    \includegraphics[width=1.15\linewidth,trim={7cm 5cm 2cm 5cm},clip]{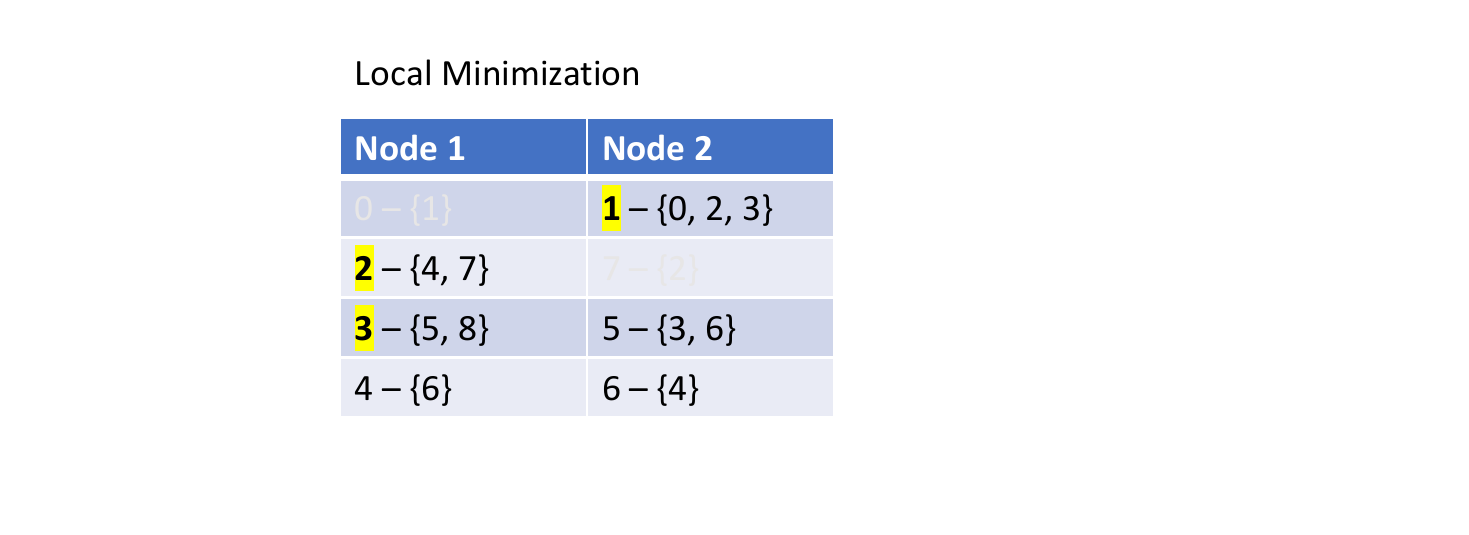}
    \caption{Vertex 7 in row 2 is removed (grayed out) as it  represents no other vertex except its same-row neighbor (namely vertex 2) and is not represented by any vertex in rows above it. Consequently, vertex 2 in row 2 is frozen (highlighted yellow) as it now represents a removed vertex (namely vertex 7).}
    \label{fig:enter-label18}
\end{figure}

\begin{figure}
    \centering
    \includegraphics[width=1.15\linewidth,trim={7cm 5cm 2cm 5cm},clip]{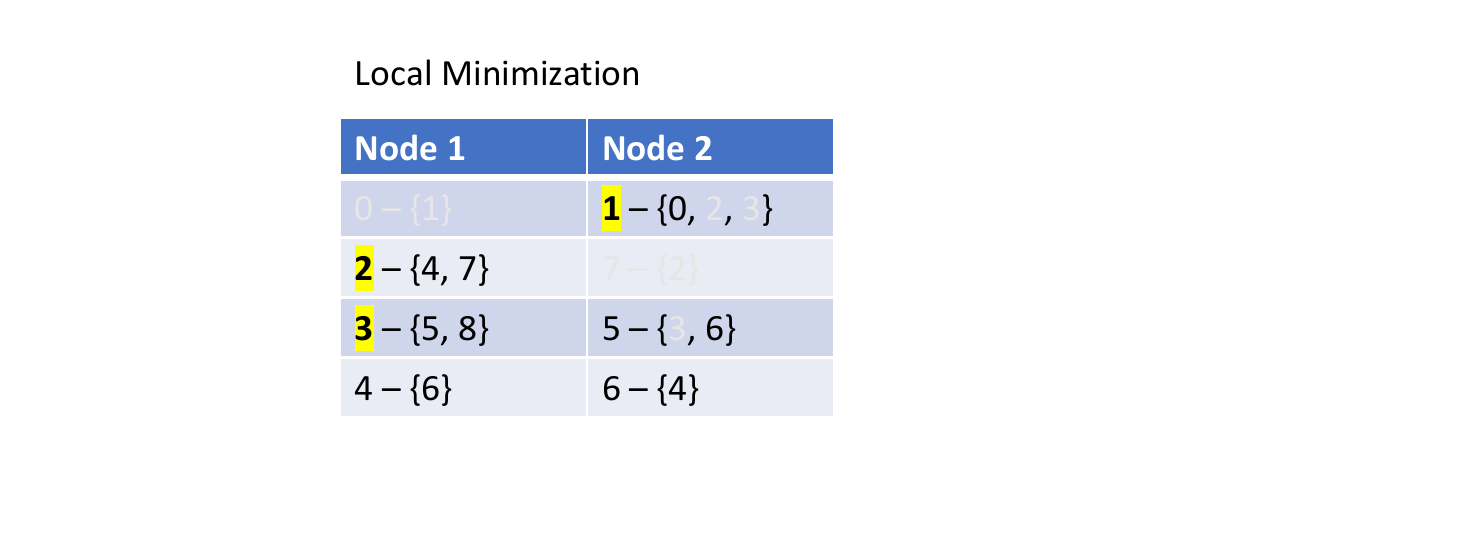}
    \caption{The frozen vertices 1, 2, and 3 are removed (grayed out) from ``represents'' list (curly brackets) of each vertex, wherever applicable. Here, vertices 2 and 3 are removed from represents list of vertex 1 and vertex 3 is removed from represents list of vertex 5.}
    \label{fig:enter-label19}
\end{figure}

\begin{figure}
    \centering
    \includegraphics[width=1.15\linewidth,trim={7cm 5cm 2cm 5cm},clip]{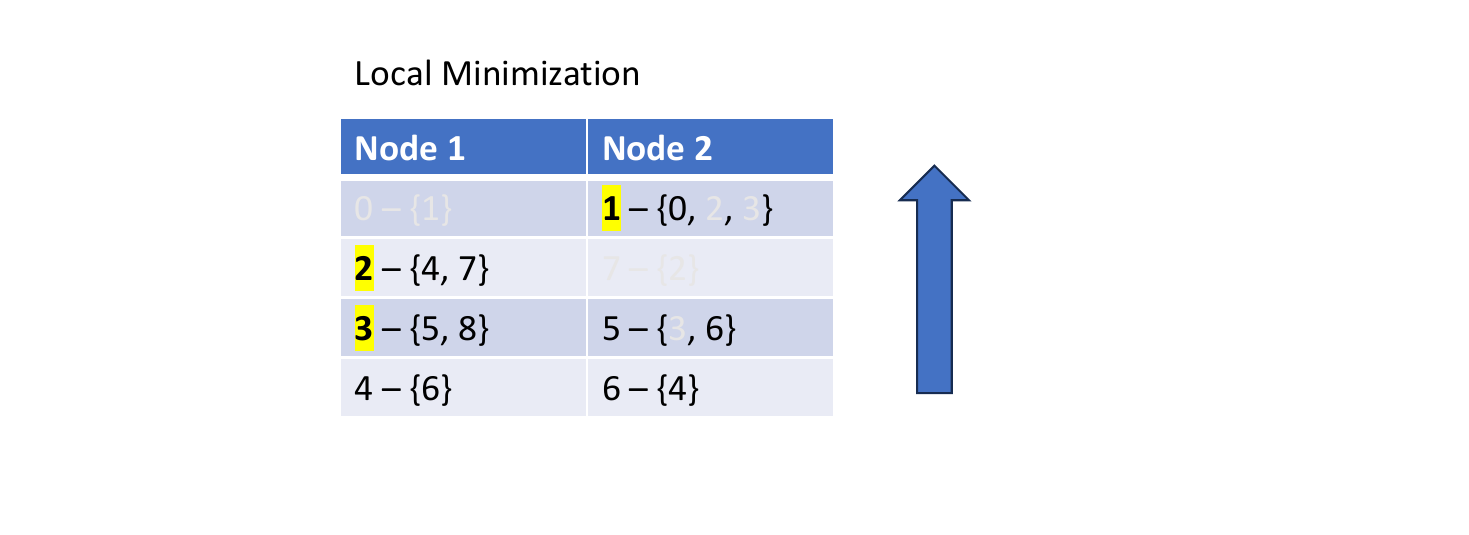}
    \caption{Arrow depicts the bottom-up elimination of vertices.}
    \label{fig:enter-label20}
\end{figure}

\begin{figure}
    \centering
    \includegraphics[width=1.15\linewidth,trim={7cm 5cm 2cm 5cm},clip]{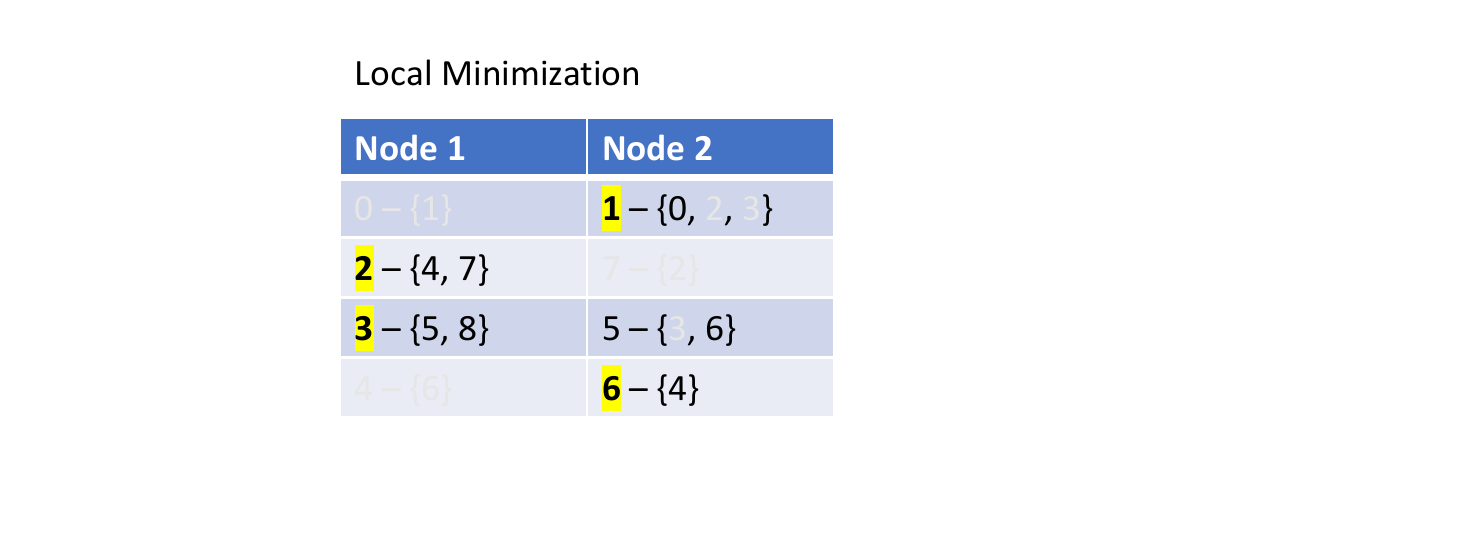}
    \caption{Start with the last row. Given that both the vertices represent only each other, we freeze one and remove the other. Specifically, vertex 4 is not represented by any vertex (or equivalently it is represented by  frozen vertex 2) and vertex 6 is represented by non-frozen vertex 5. Hence, vertex 4 is removed (grayed out) and vertex 6 is frozen (yellow highlight).}
    \label{fig:enter-label21}
\end{figure}

\begin{figure}
    \centering
    \includegraphics[width=1.15\linewidth,trim={7cm 5cm 2cm 5cm},clip]{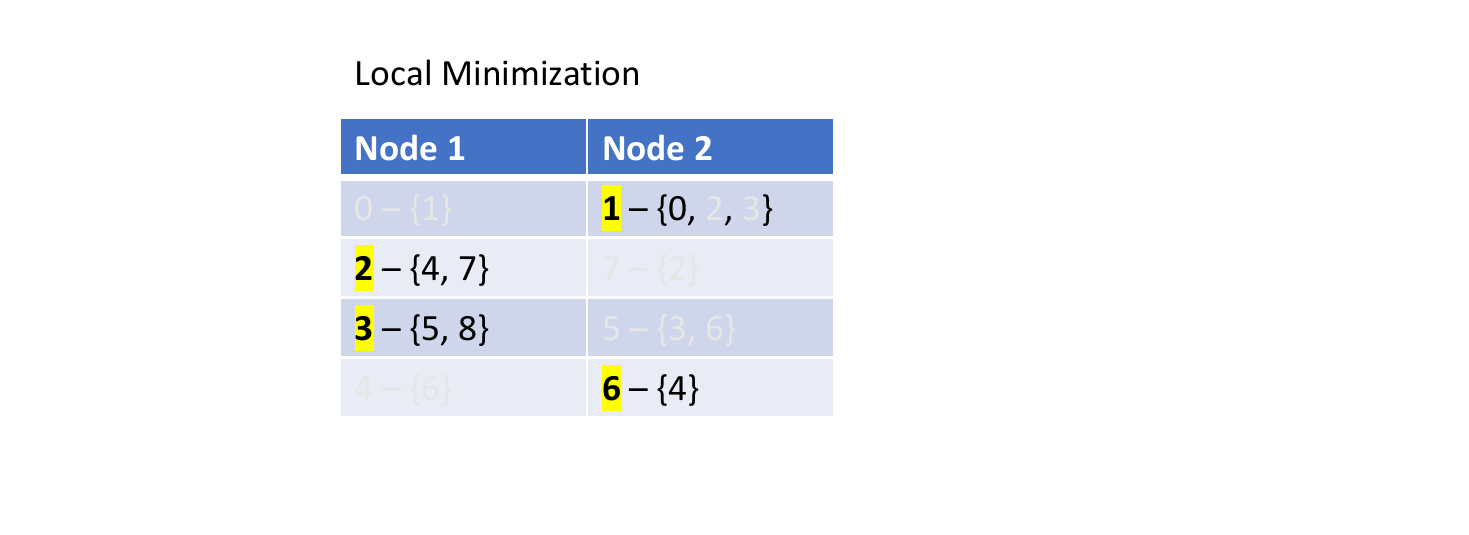}
    \caption{The frozen vertex 6 is removed (grayed out) from ``represents'' list of each vertex, wherever applicable. Here, it is removed from represents list of vertex 5. Consequently, vertex 5 does not represent any vertex. Hence, it is also removed (grayed out).}
    \label{fig:enter-label22}
\end{figure}

\begin{figure}
    \centering
    \includegraphics[width=1.15\linewidth,trim={7cm 5cm 2cm 5cm},clip]{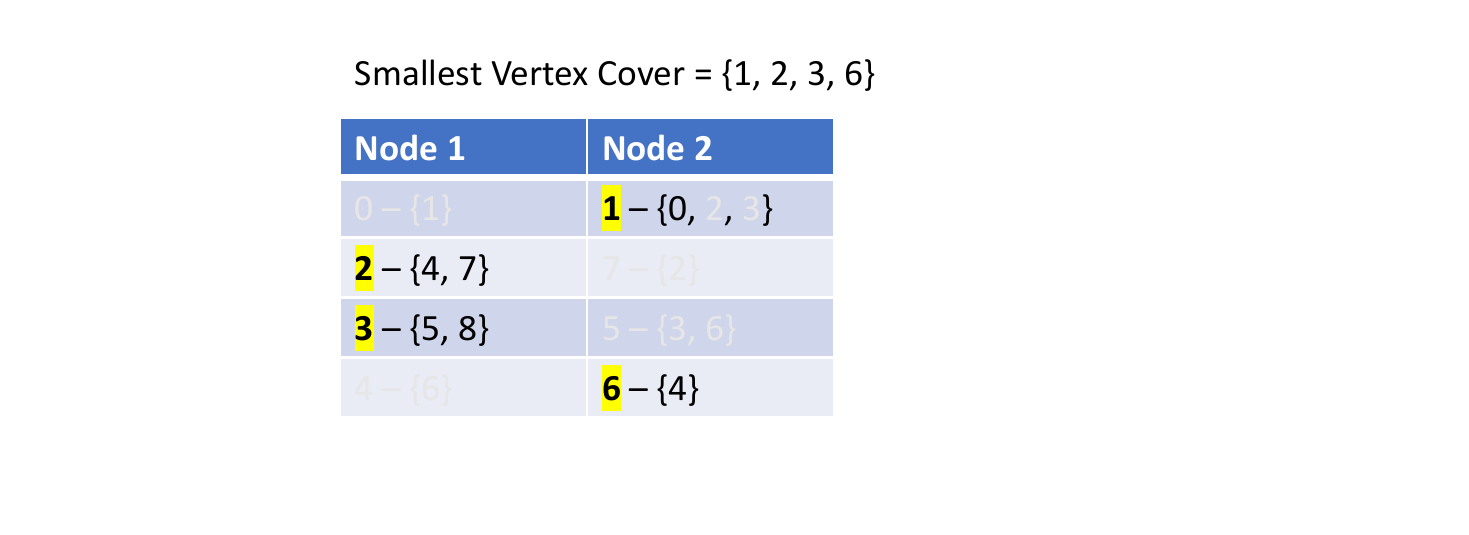}
    \caption{Only frozen vertices remain in the table. The local minimization phase terminates. The frozen vertices form the smallest vertex cover for the iteration of the algorithm whose BFS is seeded on vertex `0'.}
    \label{fig:enter-label23}
\end{figure}

\begin{figure}
    \centering
    \includegraphics[width=1.15\linewidth,trim={7cm 5cm 2cm 5cm},clip]{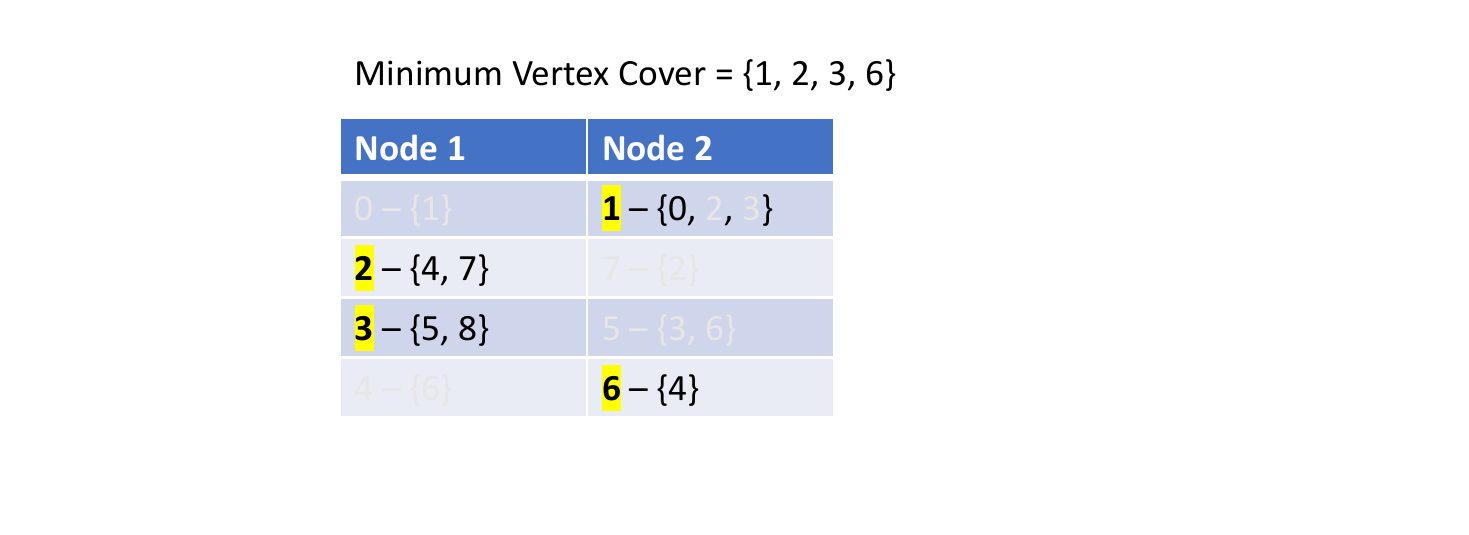}
    \caption{The size of the smallest vertex cover (= 4) is equivalent to the size of maximum matching. Hence, the smallest vertex cover is indeed the minimum vertex cover and the algorithm terminates early.}
    \label{fig:enter-label24}
\end{figure}

\begin{figure}
    \centering
    \includegraphics[width=1.1\linewidth,trim={0cm 0cm 0cm 2.12cm},clip]{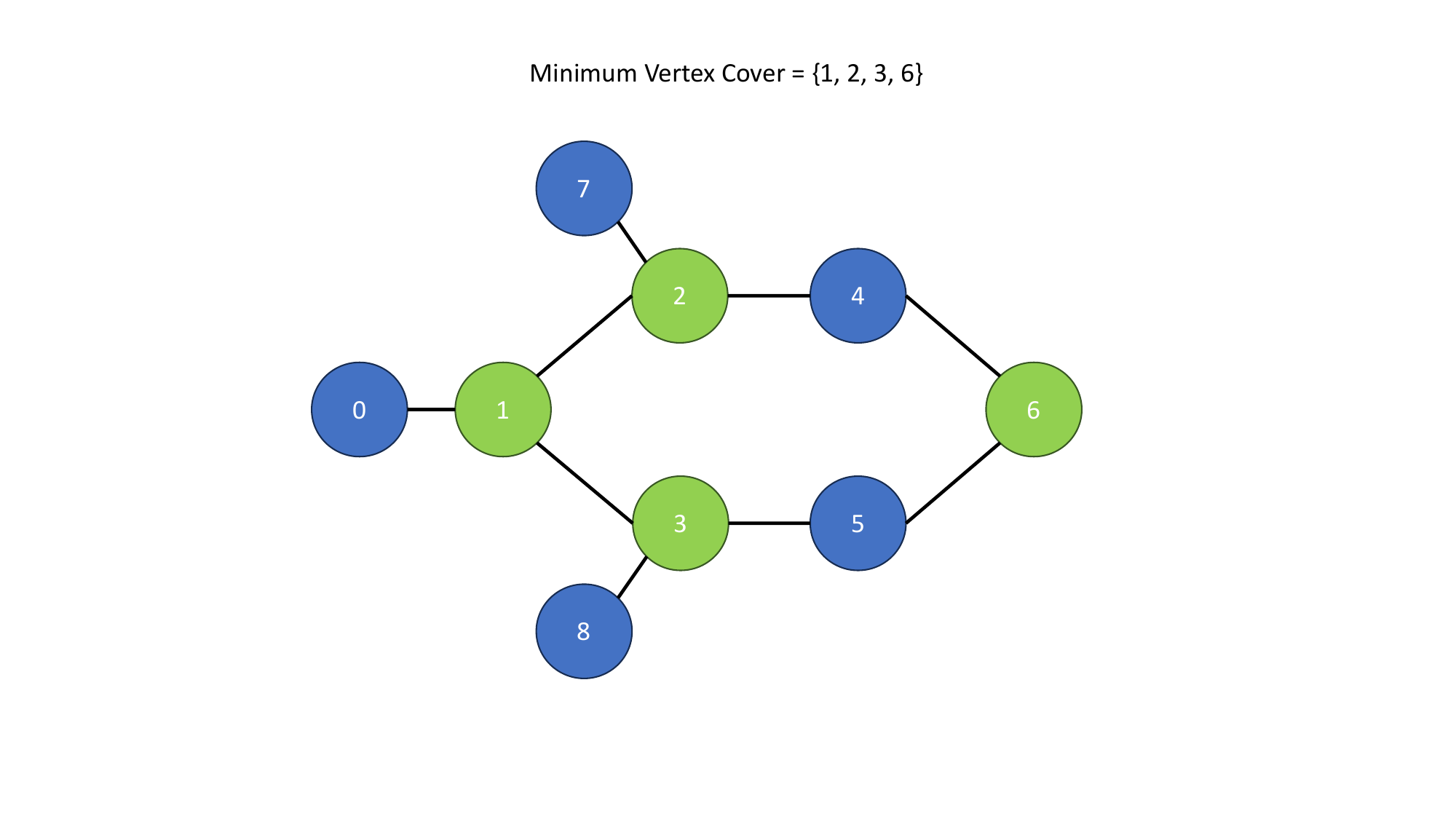}
    \caption{Vertices \{1, 2, 3, 6\} form the minimum vertex cover of size 4.}
    \vspace*{4.0625in}
    \label{fig:enter-labelLast}
\end{figure}

\begin{example}\label{eg:algoImplementation}
Consider the graph $G$ shown in Figure~\ref{fig:enter-label1}. An instance of the VC problem consists of the graph $G$ and an integer $k$ = 4. The algorithm traverses through the graph as depicted from Figure~\ref{fig:enter-label2} to Figure~\ref{fig:enter-labelLast}. The algorithm returns ``YES'' as the minimum size vertex cover shown in Figure~\ref{fig:enter-labelLast} is of size 4. 
\end{example}

% \begin{figure}
%     \centering
%     \includegraphics[width=1\linewidth]{figures/overleaf-example_8-8.pdf}
%     \caption{Enter Caption}
%     \label{fig:enter-label}
% \end{figure}

% \begin{figure}
%     \centering
%     \includegraphics[width=1\linewidth]{figures/overleaf-example_10-10.pdf}
%     \caption{}
%     \label{fig:enter-label5}
% \end{figure}

% \begin{figure}
%     \centering
%     \includegraphics[width=1\linewidth]{figures/overleaf-example_14-14.pdf}
%     \caption{Enter Caption}
%     \label{fig:enter-label}
% \end{figure}

% \begin{figure}
%     \centering
%     \includegraphics[width=1\linewidth]{figures/overleaf-example_17-17.pdf}
%     \caption{Enter Caption}
%     \label{fig:enter-label}
% \end{figure}

% \begin{figure}
%     \centering
%     \includegraphics[width=1\linewidth]{figures/overleaf-example_20-20.pdf}
%     \caption{Enter Caption}
%     \label{fig:enter-label}
% \end{figure}

\end{document}